\documentclass[twoside,12pt]{article}
\pdfoutput=1
\usepackage{graphicx}
\usepackage{epsfig}
\usepackage[utf8]{inputenc}
\usepackage{url}
\usepackage[cal=esstix]{mathalfa}
\usepackage{amsmath}
\usepackage{amsthm}
\usepackage{amssymb} 
\usepackage{float, placeins}
\usepackage{caption, subcaption}
\usepackage{dblfloatfix,fixltx2e}
\usepackage[font={it}]{caption}
\usepackage[margin=2.0cm]{geometry}
\usepackage{cancel}
\usepackage{xfrac}
\usepackage{xspace}
\usepackage{hyperref}
\usepackage{booktabs}
\usepackage[binary-units=true]{siunitx}
\usepackage{accents}
\usepackage{titling}
\usepackage[super]{nth}
\usepackage[toc,page]{appendix}
\usepackage{algorithm}
\usepackage{algpseudocode}
\usepackage{bm} 
\usepackage{authblk}

\newcommand{\tev}{\tera\electronvolt}
\newcommand{\gev}{\giga\electronvolt}

\newcommand{\mm}{\milli\meter}
\newcommand{\kNN}{\ensuremath{k\text{NN}}\xspace}

\begin{document}

\title{Deep Regression of Muon Energy with a K-Nearest Neighbor Algorithm}

\author[1]{Tommaso Dorigo}
\author[2]{Sofia Guglielmini}
\author[3]{Jan Kieseler}
\author[4]{Lukas Layer}
\author[5]{Giles C. Strong}

\affil[1,4,5]{INFN, Sezione di Padova}
\affil[2]{University of Padova}
\affil[3]{CERN}
\affil[4]{Università di Napoli ``Federico II"}

\date{\today}

\maketitle
\begin{abstract}
Within the context of studies for novel measurement solutions for future particle physics experiments, we developed a performant kNN-based regressor to infer the energy of highly-relativistic muons from the pattern of their radiation losses in a dense and granular calorimeter.
The regressor is based on a pool of weak kNN learners, which learn by adapting weights and biases to each training event through stochastic gradient descent. The effective number of parameters optimized by the procedure is in the 60 millions range, thus comparable to that of large deep learning architectures. We test the performance of the regressor on the considered application by comparing it to that of several machine learning algorithms, showing comparable accuracy to that achieved by boosted decision trees and neural networks.

\end{abstract}

\clearpage
\section{Introduction \label{s:introduction}}


The excellent generalization properties of deep Neural Networks (NN) produced a paradigm shift in the field of data analysis over the past 15 years, when significant performance gains over traditional statistical learning techniques were observed in a number of problems. Those gains came from the introduction of a very large number of free parameters in the networks architecture. Whilst this is simple to achieve (widening and deepening their layer structure), and despite tuning methods such as back-propagation~\cite{Linnainmaa_70, Werbos_81, backprop} and layer-wise pre-training ({\em e.g.} Refs.~\cite{jurgen_pretrain,hinton_pretrain}), the reliable training of such networks was only made possible thanks to recent advancements of several fundamental aspects of NNs: the development of more suitable initialisation schemes for parameters ({\em e.g.} Glorot~\cite{Glorot} and He~\cite{He}); the use of less problematic activation functions ({\em e.g.} rectified linear unit~\cite{relu}); and improvements to the gradient descent optimisation through the use of adaptive learning rates and momentum ~\cite{adagrad, rmsprop, adam}.

Prior to the above mentioned advancements, good performance could be achieved through the use of ensembling of many weak-learners, such as classification and regression trees~\cite{cart}. Two common practices in this area are represented by Random Forests (RF)~\cite{featsample, random_forests}, which decorrelate trees through the use of bootstrap resampling of training data, and feature sub-sampling; and Boosted Decision Trees (BDT)~\cite{boost, boost2}, which sequentially grow new trees to correct the errors in the prediction of the previous trees.

While in the current paradigm of deep learning and differentiable programming the use of DNNs has become widespread, there are still some domains in which they are not always the most optimal approach. Applications on tabular data are one such possible area. Here variables have strict semantic meanings tied closely to the dataset and application, are often high-level, and have no inherent ordering. These aspects limit the use of transfer learning ({\em i.e.} it is not always possible or beneficial to pre-train a DNN on a different dataset and fine-tune it to the current one), and the fact that such datasets are often comparatively small limits the degree to which randomly initialised DNNs can be trained.

Nevertheless, there have been recent developments in terms of specific architectures for dealing with tabular data, beyond basic fully connected DNNs, such as: using 1D-CNNs with a learnable feature reordering~\cite{1d_cnn}; TabNet~\cite{tabnet}, which also aims to provide a more interpretable model for use in {\em e.g.} predictive analytics; and categorical feature-embeddings~\cite{cat_embed}, in which rich, yet compact, matrix representations of categorical data are learnt, and can potentially be transferred between applications or used to produce high-level representations of features for subsequent input to classical machine learning algorithms.


Despite these advances in DNNs for tabular data, it is not unusual for winning solutions in data science competitions on tabular data to ensemble DNNs with less data-intensive models (see again, {\em e.g.}, Ref.~\cite{1d_cnn}). The flexibility of NNs and BDTs, however, can in principle be injected also in k-Nearest Neighbor (kNN) algorithms. Indeed, the idea of boosting applied to kNN is not new, and it produced demonstrations of good performance in some use cases~\cite{boostedknn1,boostedknn2,boostedknn3}. Such an approach has the potential to either provide an intermediate step, in terms of dataset-size requirements, between RFs/BDTs and DNNs, or at least a further model to include in multi-algorithm ensembles. In this work we build on those studies and include a few ideas we developed over the past two decades when applying kNN methods to particle physics applications~\cite{cmsbbb}, with the purpose of assessing their relative value with respect to today's standards for regression applications. 

\subsection {Muon energy measurements}

The occasion for this study is given by an interesting question arising in the context of optimization studies of future detectors for particle colliders that are still in the design phase, namely the measurement of the energy of highly-relativistic muons. Muons have a mass 200 times higher than that of electrons, hence they lose little energy by electromagnetic radiation when traversing dense media, behaving as minimum ionizing particles in a wide range of energies, where they are easily distinguishable from long-lived light hadrons which undergo strong interaction with atomic nuclei. These unique properties make muons excellent probes of new phenomena in particle physics: they were involved in the discovery of heavy quarks~\cite{richter,lederman,topdisc}, weak bosons~\cite{rubbia} and the Higgs boson~\cite{higgsatlas,higgscms}. Recent evidence from CMS ~\cite{hmmcms} also confirms their important role both in searches and in measurements of standard model parameters. Furthermore, the decay to final states that include muons or electrons is the preferred --when not the only-- channel for the discovery of a number of heavy particles predicted by new-physics models; at high energy, muons are usually easier to identify than electrons. This assures that muons will remain of high value for new physics searches in future high-energy colliders.

Unfortunately, the mentioned properties of muons also make the precise measurement of their energy particularly difficult. Muon energy estimates universally rely on the measurement of its momentum through precise tracking over high magnetic field integrals. The relative resolution of muon transverse momentum achieved in the state-of-the-art detectors at ATLAS and CMS ranges from 8 to 20\% (ATLAS) or 6 to 17\% (CMS) at \SI{1}{\tev}~\cite{atlasmuon,cmsmuon}, depending on detection and reconstruction details. 

In truth, muons do not behave as minimum-ionizing particles at high energy: rather, they show a relativistic rise of the energy loss~\cite{pdg} above roughly \SI{100}{\gev}. The effect is however very small in absolute terms. For that reason, in collider physics applications it has never been relied upon. However, a sufficiently thick and fine-grained calorimeter may detect the low-energy photons from muon radiation. We demonstrated in~\cite{cnnpaper} how that information, processed by deep learning algorithms, allows a $20\%$ resolution up to energies of 4 TeV. Here, we wish to use that result as a benchmark to see how far can a simple kNN algorithm be pushed on a difficult regression task, by applying techniques leveraging machine-learning ideas.

In \autoref{s:knn} below we discuss the algorithm we have developed specifically to tackle the above regression problem, in its generalities. In \autoref{s:detector} we briefly describe the idealized calorimeter we have employed for this study. In \autoref{s:features} we proceed to describe the high-level features we extract from the energetic and spatial information of each muon interaction event; these features are used as an input for the regression task. In \autoref{s:pruning} we discuss how we reduce the dimensionality of the problem, pruning the list of features of ones that appear ineffective in the regression task in combination with the others. In \autoref{s:regressor} we discuss the specific setup for the application of the algorithm to the muon energy regression, and detail our results. We offer some concluding remarks in \autoref{s:conclusion}.

\clearpage
\section{Deep regression with a kNN algorithm \label{s:knn}}

The nearest-neighbor technique is a well-known and studied technique in statistical learning theory. In its essence, the method consists in constructing estimators by averaging the properties of training events of similar characteristics to those of a test event to be classified, or whose properties need to be inferred. Similarity is gauged by the distance in a multi-dimensional space of the relevant descriptive features of the events. When speaking of kNN algorithms, the letter $k$ indicated in general that the number of training events is larger than one; $k$ then acts as a regularizer between variance and bias, as larger values reduce variance but increase bias-- the averaged events become less similar to the test point.

The performance of kNN algorithms is usually limited by the curse of dimensionality, when finiteness of training data prevent the meaningful use of a multi-dimensional distance in high-dimensional feature space. Another limitation of kNN algorithms is their limited capability of being sensitive to interdependence between the features, as the means of capturing that information relies on a single-dimensional representation, the generalized distance in the space. 

\subsection {Addressing the curse of dimensionality}

One way to address the problem introduced by large dimensionality $d$ of training data in a kNN application is the study of subspaces of the feature space. One may, {\em e.g.}, decide to ignore some of the features in computing the distance between a test point and the training neighbors: \par

\begin{equation}
    \Delta(i,j)^2 = \sum_{d=1}^{D} I(d) (x_i^{(d)}-x_j^{(d)})^2
\end{equation}

\noindent
where $I(d)$ is an indicator function equal to 1 for coordinates $d$ spanning the subspace, and 0 otherwise. Here we assume each of the variables $x^{(d)}$ to have been standardized to unit variance and zero mean, as it is common practice (although not mandatory). In a single-dimensional regression problem such as the one at hand, the prediction on the quantity of interest $E$ which is offered for test event $j$ can then be written as\par

\begin{equation}
    E(j) = \frac{\sum_{m=1}^{k} E(i_{\mathrm{kNN}}(m))}{k}
\end{equation}

\noindent
where $i_{kNN}(m)$ indicates the index of the $m$-th closest training event to $j$, according to the value of $\Delta(m,j)$.

The definition of $\Delta(i,j)$ above turns the curse of dimensionality problem into the one of locating advantageous subspaces of the feature space through a definition of the indicator function $I$, but in general the reduction of dimensionality always entails a loss of useful information. A way to reduce this loss is to consider a pool of $N_{wl}$ ``weak learners", each performing a kNN average in a different subspace of the original feature space. Each of these learners is then characterized by a different indicator function $I_{wl}$. The decomposition may be exploited by either considering a generalized distance constructed with the $N_{wl}$ ones, or by going all the way toward constructing $N_{wl}$ (partly) independent regressors, whose predictions are finally combined:\par

\begin{equation}
    E(j) = f(E_{1}(j),...,E_{N_{wl}}(j));    
\end{equation}

\noindent
the function $f$ above may be a simple average, or a weighted one in case we wish to exploit available information on the different performance of weak learners. In this application we employ a linear combination of the predictions $E(j)=\sum_{i=1}^{N_{wl}} W_{wl}(i) E_{i}(j)$ of the $N_{wl}$ weak learners.  The learner Weights $W_{wl}(i)$ are optimized by gradient descent as described {\em infra}. In the following sections, the term ``regressor" will refer to the learner created by combining the weak learner predictions, unless otherwise specified.

The decomposition of the original feature space into a number $N_{wl}$ of partially-independent subspaces of course does not guarantee against the loss of useful information for the task at hand, but the introduced flexibility lends itself to an optimization which may recover at least a part of it. In Sec.~\ref{s:algorithm} we discuss this problem and its possible solutions, as well as the one we adopted for the present work.

\subsection{Overparametrization}


It can be argued that the power of machine learning algorithms lies in their ability to be universal approximators~\cite{uni-approx} combined with relatively efficient fitting processes. In general, their modelling capabilities are improved by an overparametrization, in which the number of free parameters, weak learners, etc.\ exceed the number of parameters required to reasonably model the target function. Due to finite training datasets, however, there can be a trade-off between accurate modelling of the function as seen in the training dataset, and generalisation to unseen data.

To inject overparametrization in a kNN algorithm one must rely on training data themselves, as in essence the algorithm simply performs a local averaging of training data properties. In a regression task where the target is single-dimensional, as in the use case of our interest in this work, each training event affects the prediction by its position in the space, which is fixed, and by the value of the quantity to be estimated, which is also fixed. One way to insert flexibility in those inputs is to bias the value of the quantity to be regressed, and alter the overall weight of the event in our averaging task. This can be done by introducing two sets of parameters $b(wl,i)$ $w(wl,i)$  for each of the weak learners we have defined above, as follows:\par

\begin{equation}
    E_{wl}(j) = \frac{\sum_{m=1}^{k} E_{wl}(i_{\mathrm{kNN},wl}(m)) w(wl,m)} {\sum_{m=1}^{k} w(wl,m)} + \sum_{m=1}^{k} b(wl,m)).
\end{equation}

\noindent
The introduction of a weight and a bias for every training event, independently for each weak learner, provides a large number of parameters to the regression task. In our application to muon energy regression it corresponds to a total in the range of ${\cal{O}}(10^7-10^8)$ free parameters. 

\subsection{Gradient descent}

Once one defines a loss function for the problem at hand, which in our case is a function of true and predicted energy of each muon in the batch of data under evaluation, it is straightforward to compute derivatives of the loss with respect to the weights and biases defined above. In our implementation of the algorithm we did not rely on automated differentiation methods, as an exact solution can also be obtained manually within a couple of pages of a notebook.

The need to optimize weights and biases of the overparametrized weak learners by stochastic gradient descent, however, requires some discussion. As the kNN algorithm relies for the prediction on a set of training data, the data required to perform batch gradient descent need to come from an independent sample. We thus have to split our training data sample into a prediction and a training set; the first contains events that enter the averaging and the actual predictions $P=E(j)$ as above, and the second contains the data used to learn optimal values of the parameters. In Sec.~\ref{s:algorithm} below we discuss how the data are handled for this task.

\subsection {Loss function \label{s:loss}}

The most straightforward choice for a loss function in a linear regression task is the mean squared error (MSE):  \par

\begin{equation}
    \alpha_0 \sum_{i=1}^{N_{batch}} \frac{(T_i-P_i)^2}{\sigma_0^2}
\end{equation}

\noindent
for $N_{batch}$ events of true energies $T_i$ and predictions $P_i$. The denominator $\sigma_0^2$ may be used as a constant scaling factor or to introduce a dependence on the resolution we target for different true energies, if we set $\sigma = \sigma(T_i)$. For example, a calorimeter-inspired loss would suggest the choice $\sigma = k \sqrt{T_i}$; however, since in our application we are concerned with obtaining the best estimate of high-energy muons, we avoid such a choice, which would down-weight the importance of errors on precisely those events. 

The parameter $\alpha_0$ may be used when other terms are added to the loss (see {\em infra}). In the problem we are considering, we are faced with large non-Gaussian tails in many of the observable features of muon energy deposits and related quantities, and therefore we wish to reduce the impact of those outliers on the value of the loss. We accomplish this by further modifying version of the MSE term below:\par

\begin{equation}
    L_1 = \frac{\alpha_0}{N_{batch}} \sum_{i=1}^{N_{batch}} \left[1 - \exp\left(\frac{(T_i-P_i)^2}{2 \sigma_0^2}\right)\right].
\end{equation}

The importance of the denominator in the exponential term now has increased, as it is no longer an overall scaling, albeit one depending on true muon energy: the effect of outliers on the loss now depends on its value. By choosing it carefully it is possible now to focus on the range of predictions which we care to regress, leaving outliers to their fate. 

One specific problem of the use case we have considered in this work is the existence of a persistent bias in the predictions that can be obtained by kNN regression. Being confronted with predicting an energy value by averaging  training data of true energy in a fixed range, the regressor's output is naturally pulled toward the mean of the range, and thus biased to overestimate small energies and underestimate high energies. While corrections can be applied to the regression output, these are observed to spoil the overall performance of the regression, as they act after the optimization task and the minimization of the loss. We therefore designed a penalization factor in the loss function to reduce the bias during optimization, avoiding the misalignment of a two-step procedure with the goal of the regression.

If we consider the optimal regressor as one which produces a prediction $P$ that carries no bias for any true value $T$ of the target, this may be written by the condition \par

\begin{equation}
    {\mathbb{E}}(P) = T,
\end{equation}

\noindent
where with ${\mathbb{E}}$ we indicate an expectation value. In practical terms, we can impose this condition on a set of events of true energy $T$ by asking that\par

\begin{equation}
    \hat{P} = T,
\end{equation}

\noindent
where $\hat{P}$ is the average of the predictions for events of true energy $T$. We may enforce this condition on our predictors, through the optimization of weights and biases, if we bin our data in a number $N_{T}$ of bins in the interval of interest of true energy values, and penalize the loss function by the introduction of the following term:\par

\begin{equation}
    L_2 = \alpha_1 \sum_{m=1}^{N_T} \sum_{n=1}^{N_T} [\hat{P_{m}} - \hat{P_{n}} - T_m + T_n]^2 / \sigma(m,n)^2
\label{eq:L1}
\end{equation}

\noindent
where $\alpha_1$ is a hyperparameter determining the relative strength of the penalization, and the numerator in the sum corresponds to the observed minus estimated difference in mean values between two bins $m$ and $n$, $\Delta_{mn,obs}-\Delta_{mn,exp}$. 

The denominator in Eq.~\ref{eq:L1}, $\sigma(m,n)^2$, requires a separate discussion. In general, one would like this term to modulate the weight of different discrepancies between differences of mean values and expected differences, providing a linearity constraints toward the target $\hat{P}=T$. We may arrive at a suitable formulation by considering that an individual prediction $P$, derived from calorimetric measurements of energy deposits, usually carries an uncertainty that scales with its square root, $\sigma_{P}=k \sqrt{P}$, with some value of $k$ which we take here to be unity for simplicity; this behavior is caused by the Poisson fluctuations in the development of showers in the detector. The average of the $N_m$ predictions contained in bin $m$ of the histogram of predictions for a batch of events will consequently carry an uncertainty of the order of $\sigma_{\hat{P_m}} = \sqrt{T_m/N_m}$. A back-of-the-envelope estimate of the uncertainty of the difference at the numerator in the sum of $L_2$ above can then be written as

\begin{equation}
    \sigma_{\Delta_{mn,obs}-\Delta_{mn,exp}}^2 \simeq \frac{T_m}{N_m}+ \frac{T_n}{N_n}.
\end{equation}

\noindent
The above expression could already be a meaningful definition for the scaling denominator in $L_2$ above. However, as defined so far, it would provide in the loss term equal importance to deviations from expected differences between two close bins and two bins that are wide apart along the true energy spectrum. This is not the most effective way to enforce an equalized response. If, however, we weigh each term in the double sum of Eq.~\ref{eq:L1} by the expected difference $\Delta_{mn,exp}$ we obtain the desired result: the noise contributed by large deviations in neighboring bins will be dampened with respect to the large contribution of significant variations over larger energy differences. The full expression of the denominator in the terms of the double sum in Eq.~\ref{eq:L1} is thus

\begin{equation}
    \sigma(m,n)^2 =  \frac{T_m N_n + T_n N_m}{N_m N_n (T_m-T_n) } .
\end{equation}

With the introduction of the $L_2$ factor, the loss function becomes in general non-convex, as the two requirements of minimum squared error and of linearity of mean predictions are in general in conflict -- which is self-evident as the MSE term by itself converges to a biased result which shows non-linearity, as discussed above. However, by tuning the relative weight of $\alpha_0$ and $\alpha_1$, it is possible to converge to meaningful results. The solution will not, in general, be one of minimum MSE; but it will be the one which has minimum MSE among those which manage to constrain the non-linearity to within a certain level. 

Similarly to the MSE-like term mediated by hyperparameter $\alpha_0$, the penalization term of the loss function can be differentiated easily with respect to the weights and biases, such that batch gradient descent effectively produces a loss minimization. The default loss we employ in this work is therefore the following:

\begin{equation}
    L = L_1 + L_2
\end{equation}

\noindent
We also provide here the value of $\sigma_0$ we have used in our regression task: we have chosen $\sigma_0=0.8~\mathrm{TeV}$, which corresponds to a $40\%$ relative uncertainty on 2-TeV muons, which sit in the middle of the range of muon energies we focus on. This choice, introduced in the Gaussian term of $L_1$, corresponds to focusing more on improving the prediction for events that are measured with better than $80\%$ accuracy of so, and basically accounting for outliers with a constant penalization in the loss. The value of the hyperparameters $\alpha_0$ and $\alpha_1$ are instead rather non-enlightening. They have been set respectively to 50 and $10^{-6}$, after a complicated attempt to balance the weight of each of the $L_1$, $L_2$ terms in the gradient descent.

\subsection{Other figures of merit}

The loss function constructed as described {\em supra} is a good proxy to the target of the regression task, and it fulfils the task of finding weights and biases corresponding to a good balance of MSE and linearity. However, its value on a set of test data by itself does not fully describe the overall performance of a regression. For instance, the precision of the estimates for muons of low energy is certainly less important than the precision of estimates for high-energy muons, because independent estimates of muon curvature in any realistic detector make the former much less relevant than the latter. One way to handle this would be to model our focus on higher energies in a loss that weighted the precision with the true energy of muons; however, after experimenting with loss definitions alternative to the one described {\em supra} we decided to stick with the one presented above, which remains a more general way to address the regression task.

We define alternative figures of merit to gauge the performance of the regression. One of them is obtained by extracting from regression results the root mean-squared-error of the prediction, as a function of muon true energy, and use it in combination with a hypothetical independent measurement of muon momentum based on the curvature of the track, performed upstream of the calorimeter. We choose a relative uncertainty on muon momentum 

\begin{equation}
    RMS_{tr} = \frac{\delta p}{p} = 0.2 p/\mathrm{TeV}
\label{eq:deltap}
\end{equation}

\noindent
as a baseline, a number in the same ballpark of the performance of the ATLAS detector (which features an axial field of the same intensity as the one we modeled for our simulation, 2 Tesla). A combination of the curvature-driven and radiation-driven estimates of muon energy ($E=p$) in the energy range we are considering) is obtained as 

\begin{equation}
    RMS_{comb}(E) = \sqrt{\frac{RMS_{tr}(E)^2 RMS_{cal}(E)^2}{RMS_{tr}(E)^2+RMS_{cal}(E)^2}}
\label{eq:combres}
\end{equation}

\noindent
where $RMS_{cal}(E)$ indicates the mean-squared-error of the energy prediction of the regressor, for true energy $E$. We will use the combined RMS defined above in reporting results as a function of muon energy {\em infra}. For now, we define a figure of merit of the regression task as ``maximum combined resolution":

\begin{equation}
    Max Res = \max_{E=0.05 - 4~\mathrm{TeV}} {RMS_{comb}(E)}
\end{equation}

\noindent
The function defined above is a principled summary of the ultimate capabilities of a detector which exploits calorimetry together with tracking to estimate muon energies. We note, however, that the above definition is liable to stochastic fluctuations much more than any measure of performance integrated over all the spectrum. For this reason, we will only marginally rely on it in the following.

An idea to overcome the stochasticity of $Max Res$ as defined above is to compute the integral, over the [0.05-4 TeV] range, of the difference between the tracker-only resolution function (Eq.~\ref{eq:deltap}) and the combined resolution (Eq.~\ref{eq:combres}):

\begin{equation}
    Area = 0.2532 \int_{0.05}^{4} [RMS_{tr}(E) -RMS_{comb}(E)] dE 
\end{equation}

\noindent
where the factor $1/3.95=0.2532$ serves as a normalization parameter for the investigated energy range. In our tests we found that the above figure of merit is however also not entirely satisfactory, as it is also quite sensitive to fluctuations with respect to real variations due to different performance of regressors. In our regression study we prefer to this and the previous figure of merit two others which take a different approach to the description of performance. These are constructed from the means and standard deviations of marginal distributions of kNN-predicted energy for specific true muon energy values. We select muons of true energy in the four ranges $[0.95,1.05]~\mathrm{TeV}$, $[1.95,2.05]~\mathrm{TeV}$, $[2.95,3.05]~\mathrm{TeV}$, and $[3.95,4.05]~\mathrm{TeV}$, and compute the mean $\hat{E_{p}}(E_{t})$ and RMS $\sigma_{E_p}(E_{t})$ of their predicted energy $E_p$, which we identify below by indicating the mean energy of each interval (1,2,3,4 TeV). We then compute:

\begin{equation}
    Discr_{24} = \frac{\hat{E_p}(4)-\hat{E_p}(2)}{\sqrt{\sigma_{E_p}(4)^2+\sigma_{E_p}(2)^2}}
\end{equation}

\begin{equation}
    Discr_{13} = \frac{\hat{E_p}(3)-\hat{E_p}(1)}{\sqrt{\sigma_{E_p}(3)^2+\sigma_{E_p}(1)^2}}
\end{equation}

\noindent
The above figures of merit are useful proxies of the power of hypothesis tests that try to distinguish muons of two and four TeV, or muons of 1 and 3 TeV. We found that they are quite robust, as expected given the good statistical properties of mean and SQM, and they provide valuable information of the quality of the regression task.

\subsection {Hyperball optimization \label{s:HB}}


An additional optional feature of the algorithm we developed focuses on the metric of the feature space\footnote{ This feature is the basis of the ``Hyperball algorithm", which was first used for a regression task in a background modeling regression~\cite{cmsbbb}; similar ideas were studied in~ \cite{knnhb}.}. While the subsampling of the space operated by weak learners already addresses in part the different relative importance of the variables in determining the similarity of events (by explicitly voiding a few of the space dimensions), within the subspace studied by each weak learner there still is a hierarchy that may be exploited for more accurate selection of the closest neighbors. 

If the option is turned on, the algorithm attempts to gauge the local properties of the dependence of each feature on the target, by directly estimating the bias which the prediction incurs on by relying on an average over a finite interval of each feature.

\begin{center}
\begin{figure}[h!]
\centerline{\includegraphics[width=0.6\textwidth]{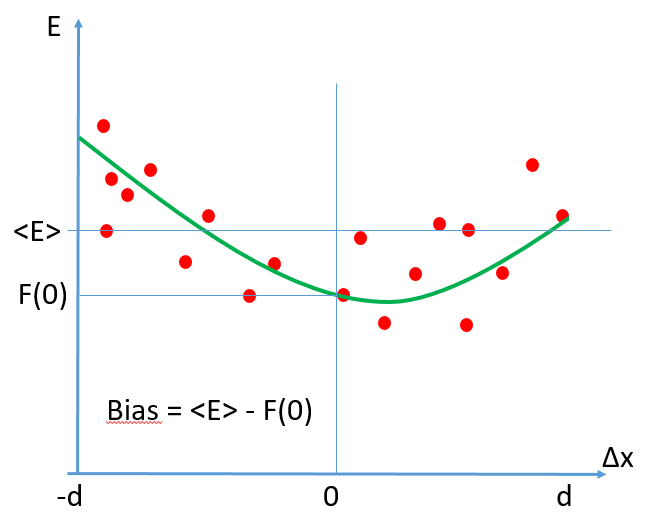}}
\caption{\em Graphical description of the bias in the calculation of the energy E from the averaging of events located at distances $\Delta x$ along one variable, if the true dependence of $E$ on $x$ is quadratic.}
\label{f:HB}
\end{figure}
\end{center}

\noindent
If we consider a one-dimensional interval of relative distances $[-d,d]$ from event $j$ along feature $V$, and a set of $j$'s $K_0$ closest neighbors, we may picture the regression to $T_j$ as the averaging of the true energy $T$ of the neighbors laying in $[-d,d]$. The average produces a unbiased prediction as long as the unknown dependence of $T$ on $V$ is linear within the interval. If the dependence is more complex, a bias will be present because the average corresponds to using an estimate of the integral of function $T(V)$ to infer the value $T(0)$ at the center of the interval. An estimate of the above bias for each feature can be produced by comparing the average of $T(V)$ in the range $[-d,d]$ with the value of the offset parameter $c$ of a quadratic function $f(\Delta V)=a \Delta V^2 + b \Delta V + c$ interpolating the observed values of $T$ as a function of their distance along $V$ from $V_j$ (see \autoref{f:HB}).

The calculation, while of course subjected to statistical imprecisions as well as systematic ones (arising, {\em e.g.}, from a more complex behavior of the true dependence than a quadratic one), allows one to derive separate estimates of the biases $b_{i}$ for each direction $V_i$ around each test point $j$ of the feature space. They may then be used to ``reshape" the hyperellipsoid that allows one to find the most relevant $K_1$ closest training events to the event for which a prediction is required. In practice, for each test point a first selection of $K_0$ neighbors is performed with an Euclidean metric; the biases are then estimated using those events, and finally used to select the $K_1$ events which are closest to the test point according to the modified metric:\par

\begin{equation}
    d = \sum_{i=1}^{N_D} \frac{1}{b_i/b_{max}+\epsilon} (x_{i,j}-y_{i,j})^2 
\end{equation}

\noindent
where $x$, $y$ are respectively the test point and a neighbor, $b_i$ is the bias for variable $i$, and $b_{max}$ is the maximum bias observed among the active dimensions of the considered subspace; $\epsilon$ is a small number inserted to prevent divergence.

A problem of the approach described above is the large CPU consumption, due to the need to determine biases for each test point by identifying a larger number $K_0$ of neighbors according to the default metric, before the hyperellipsoid can be defined around the point. In our studies of muon energy regression we have also observed that the performance gains of the procedure are not always significant, depending on a number of parameters (subspace definition, breadth of larger initial neighboring pool). For these reasons in the current preliminary version of this work we do not present results based on this additional technique, whose application is left as future work.


\section{Detector geometry and simulation}\label{s:detector}

Following the definition of the problem in ~\cite{cnnpaper}, we consider a homogeneous lead tungstate cuboid calorimeter, with a total depth in $z$ of $\SI{2032}{\mm}$, corresponding to 10 $\lambda_0$, and a spatial extent of $\SI{120}{\mm}$ in $x$ and $y$. The calorimeter is segmented in 50 layers in $z$, each with a thickness of $\SI{39.6}{\mm}$, which corresponds to 4.5 radiation lengths, such that electromagnetic showers are well resolvable. A layer is further segmented in $x$ and $y$ in $32 \times 32$ cells, with a size of $\SI{3.73}{\mm} \times \SI{3.73}{\mm}$. This results in 51,200 channels in total.

\begin{figure}[h!]
\centering
\includegraphics[width=0.7\textwidth]{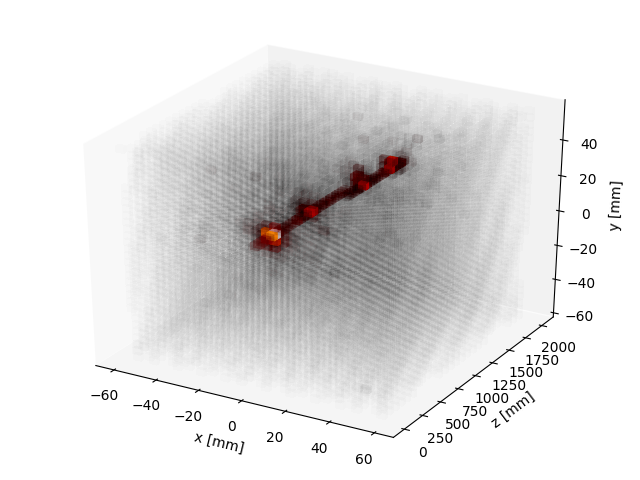}
\caption {\em Event display of a 290 GeV muon traversing the simulated calorimeter along the $z$ direction. The colours indicate the intensity of energy deposits. Black corresponds to zero, red to intermediate, and white to high energy.}
\label{f:geometry}
\end{figure}

\subsection {Datasets}

We generate muons  with a momentum $P = P_z$ in the $z$ direction, of magnitude $|P|$ between $\SI{50}{\gev}$ and $\SI{8}{\tev}$. This choice of range extends beyond the conceivable momentum range of muons produced by a future high-energy electron-positron collider such as CepC or FCC-ee, thus enabling a unbiased study of the measurement of that quantity in an experimentally interesting scenario. For muons of up to 2-3 TeV one could in principle still conceive a curvature-based estimate, but the construction demands of the powerful magnetic fields necessary for that task would seriously constrain the space of detector design choices; hence our interest in exploring the calorimetric alternative.

The generated initial muon position in the $z$ coordinate is set to $z=-\SI{50}{\mm}$ with respect to the calorimeter front face; its $x$ and $y$ coordinates are randomly chosen within $|x|\leq\SI{20}{\mm}$ and $|y|\leq\SI{20}{\mm}$. The momentum components in $x$ and $y$ direction are set to zero. The interaction of the muons with the detector material is simulated using Geant4~\cite{GEANT4}.  In total a set of 887,000 events are generated, out of which 187,000 are set aside to gauge the regression performance. The remaining 700,000 are used to train the regressor. A further set of 265,000 events are generated at fixed values of muon energy (E = 100, 500, 900, 1300, 1700, 2100, 2500, 2900, 3300, 3700, 4100, 4500, 4900 GeV), to test the performance of the regression on mono-energetic samples. A copy of these datasets is publicly available in Ref.~\cite{dataset}.

In order to consider a realistic scenario, and meaningfully compare the curvature-based and calorimetric measurements, we assume that the calorimeter is immersed in a constant $B=2T$ magnetic field, oriented along the positive $y$ direction.  
The detector geometry, as well as a possible radiation pattern of a muon entering the calorimeter is shown in \autoref{f:geometry}. Even at a relatively low energy of $\SI{290}{\gev}$, the radiation pattern is clearly visible.

\section{Event features \label{s:features}}

The regression task we set up in \autoref{s:regressor} employs 28 event features extracted from the spatial and energy information collected in the calorimeter cells, using domain knowledge to aggregate the inputs in an informative way. The 28 features are described in detail in ~\cite{cnnpaper}. Some of the features describe general properties of the energy deposition ({\em e.g.}, the sum of all deposited energy above or below a $E=\SI{0.1}{\gev}$ threshold). Some of them are partly reliant on fine-grained information (moments of the energy distribution, in five regions of detector depth: $z<\SI{400}{\mm}$, $400<z<\SI{800}{\mm}$, $800<z<\SI{1200}{\mm}$, $1200<z<\SI{1600}{\mm}$, and $z>\SI{1600}{\mm}$; and imbalance of the deposited energy in the transverse plane). A few more variables are instead computed from a full-resolution clustering of the energy deposition pattern (see ~\cite{cnnpaper}). One final variable is specifically constructed to estimate the curvature of muons of momenta below $\simeq 500 GeV$ (as above that value there is almost no extractable information) by detecting the spreading of the radiation pattern along the $x$ coordinate caused by the small bending along $x$ that muons follow as they penetrate in the calorimeter.

        \begin{itemize}
            \item V[0]: The total energy recorded in the calorimeter in cells above the $E_{\text{thr}}>\SI{0.1}{\gev}$ threshold;
            \item V[1]: We define $H_x = \sum_i E_i\cdot\Delta x_i$ and $H_y = \sum_i E_i\cdot\Delta y_i$, where $\Delta x_i$ and $\Delta y_i$ are the spatial distances in the $x$ and $y$ directions to the centre of the cell which is hit by the muon at the calorimeter front face; from these we derive V[1] = $\sqrt{H_x^2+H_y^2}/\sum_i E_i$. In this calculation, all cells are used;
            \item V[2]: This variable results from the same calculation extracting $V[1]$, but it is performed using in all sums only towers exceeding the $E_{\text{thr}}=\SI{0.1}{\gev}$ threshold;
            \item V[3]: The second moment of the energy distribution around the muon direction in the transverse plane, computed with all towers as 
            $V[3]= \sum_i [E_i (\Delta x_i^2 + \Delta y_i^2)]/ \sum_i E_i$, 
            where indices run on all towers and the distances are computed in the transverse plane, as above;
            \item V[4]: The same as V[3], but computed only using towers located in the first \SI{400}{\mm}-thick longitudinal section of the detector along $z$;
            \item V[5]: The same as V[3], but computed only using towers in the $400<z_i<\SI{800}{\mm}$ region;
            \item V[6]: The same as V[3], but computed only using towers in the $800<z_i<\SI{1200}{\mm}$ region;
            \item V[7]: The same as V[3], but computed only using towers in the     $1200<z_i<\SI{1600}{\mm}$ region;   
            \item V[8]: The same as V[3], but computed only using towers in the $z_i\geq\SI{1600}{\mm}$ region;
            \item V[9]: The number of energy clusters seeded along the muon trajectory (type-1 clusters);
            \item V[10]: The maximum number of cells among type-1 clusters;
            \item V[11]: The maximum total energy among type-1 clusters;
            \item V[12]: The maximum extension along x of type-1 clusters;
            \item V[13]: The maximum extension along y of type-1 clusters;
            \item V[14]: The maximum extension along z of type-1 clusters;
            \item V[15]: Second-highest maximum energy in a $3\times\!3\times\!3$ cubic box from cells not included in type-1 or type-2 clusters;
            \item V[16]: Average number of cells included in type-1 clusters.
            \item V[17]: The number of clusters seeded by a cell not belonging to the muon trajectory (type-2 clusters);
            \item V[18]: The maximum number of cells among type-2 clusters;
            \item V[19]: The maximum total energy among type-2 clusters;
            \item V[20]: Ratio between maximum energy and maximum number of cells of type-2 clusters;
            \item V[21]: Average number of cells included in type-2 clusters.
            \item V[22]: The first moment of the energy distribution along the x axis, relative to the x position of the incoming muon track;
            \item V[23]: The first moment of the energy distribution along the y axis, relative to the y position of the incoming muon track;
            \item V[24]: Estimate of muon momentum extracted from a fit to the muon bending;
            \item V[25]: Maximum energy in a $3\times\!3\times\!3$ cubic box from cells not included in type-1 or type-2 clusters;
            \item V[26]: Sum of energy recorded in cells with energy below \SI{0.01}{\gev};
            \item V[27]: Sum of energy recorded in cells with energy between 0.01 and $E_{thr}=\SI{0.1}{\gev}$.  
        \end{itemize}

\noindent
Marginal distributions of the 28 features, as well as two-dimensional temperature graphs of the dependence of the features on true muon energy, are shown in Fig.~\ref{f:feats} below.

\begin{figure}[ht]
 	\begin{center}
		  \includegraphics[width=\textwidth]{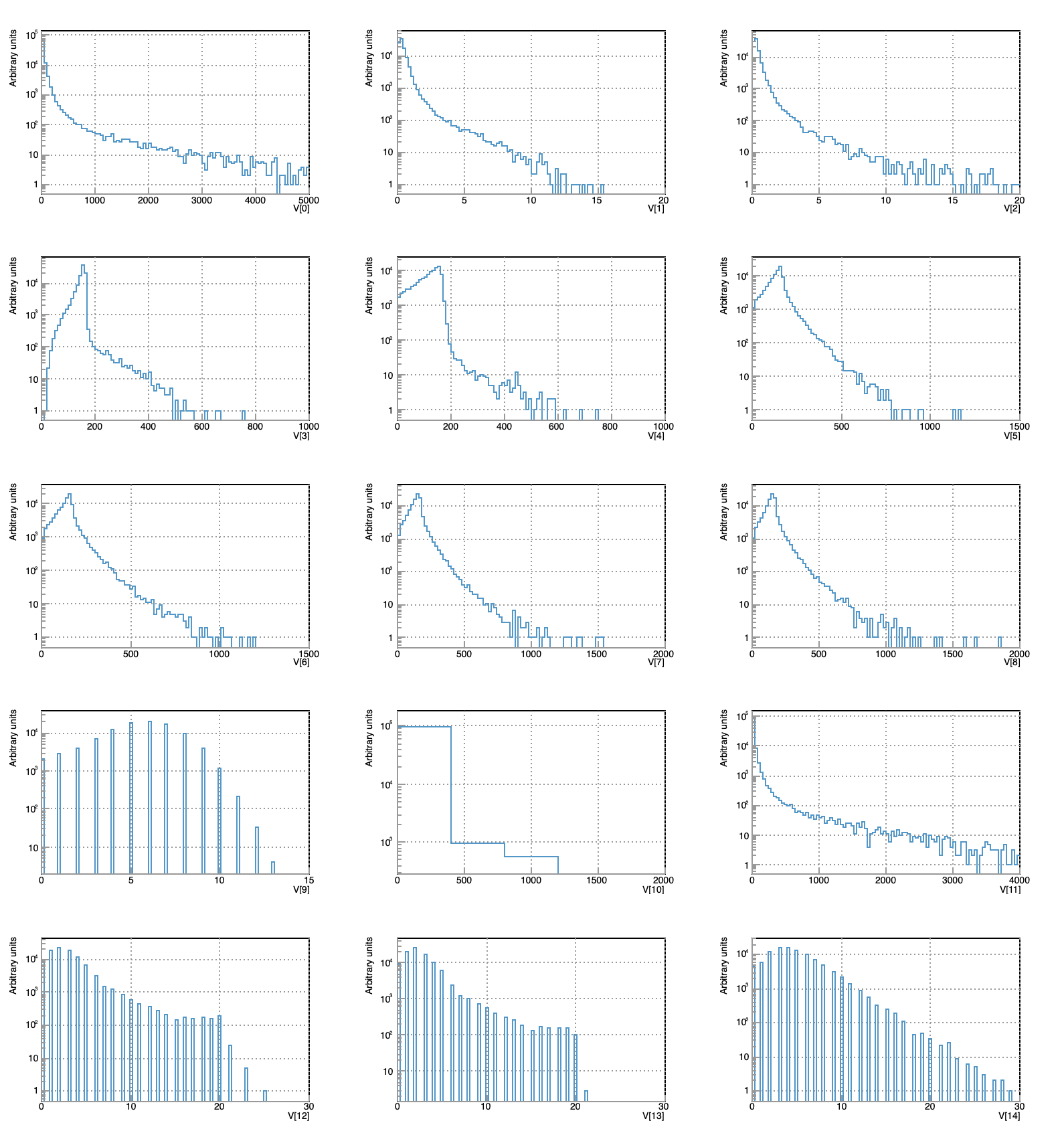}
     \caption{\em Marginals of event features. Top row: V[0],V[1], V[2]; second row: V[3],V[4], V[5]; third row: V[6],V[7],V[8]; fourth row: V[9],V[10],V[11]; bottom row: V[12],V[13],V[14]. Features are defined in Section~\ref{s:features}.}
        \label{f:feats}
    \end{center}
\end{figure}

\begin{figure}[ht]
 	\begin{center}
		  \includegraphics[width=\textwidth]{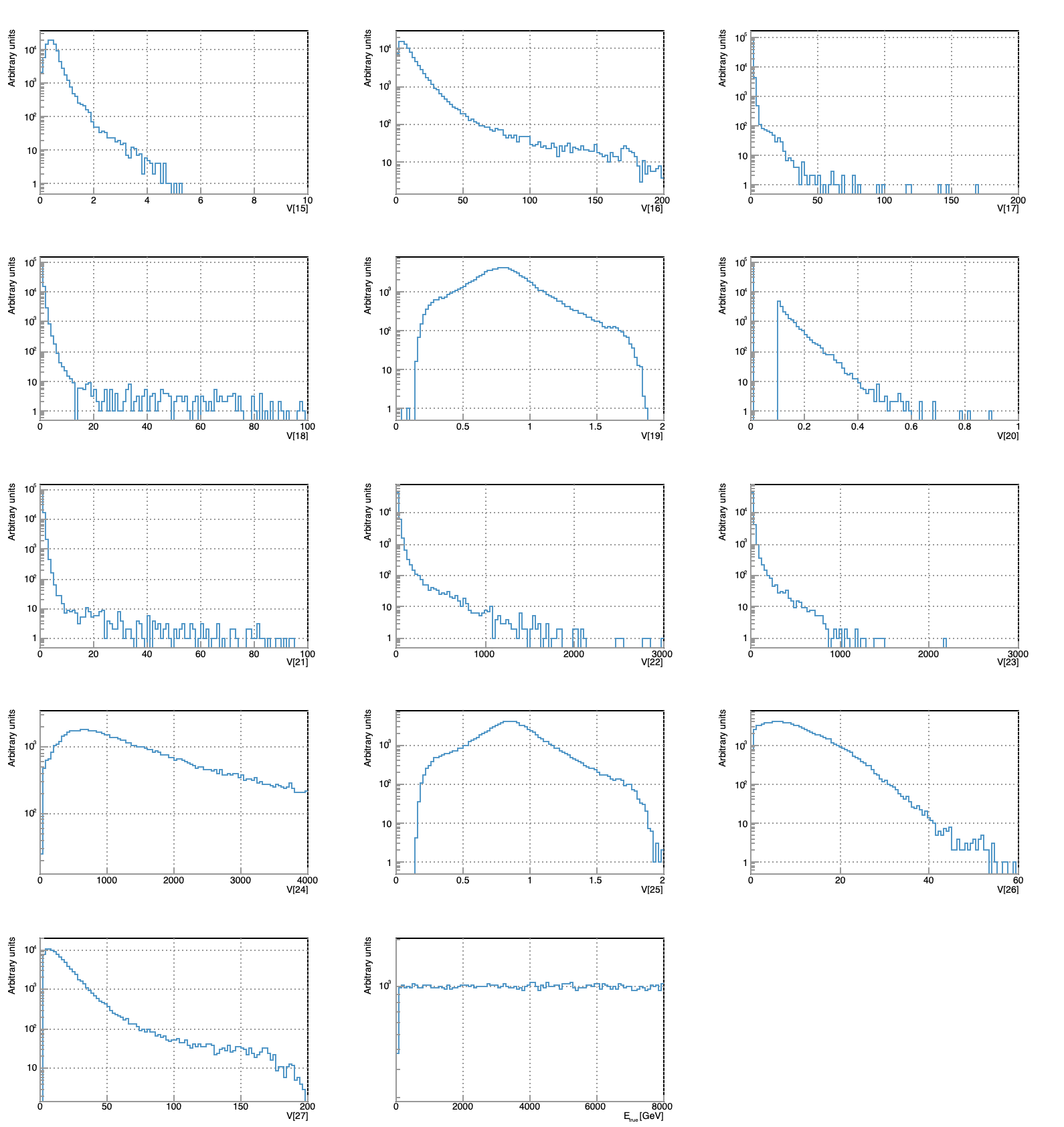}
     \caption{\em Marginals of event features and true muon energy. Top row: V[15],V[16], V[17]; second row: V[18],V[19], V[20]; third row: V[21],V[22],V[23]; fourth row: V[24],V[25],V[26]; bottom row: V[27], true muon energy. Features are defined in Section~\ref{s:features}.}
        \label{f:feats}
    \end{center}
\end{figure}

\begin{figure}[ht]
 	\begin{center}
		  \includegraphics[width=\textwidth]{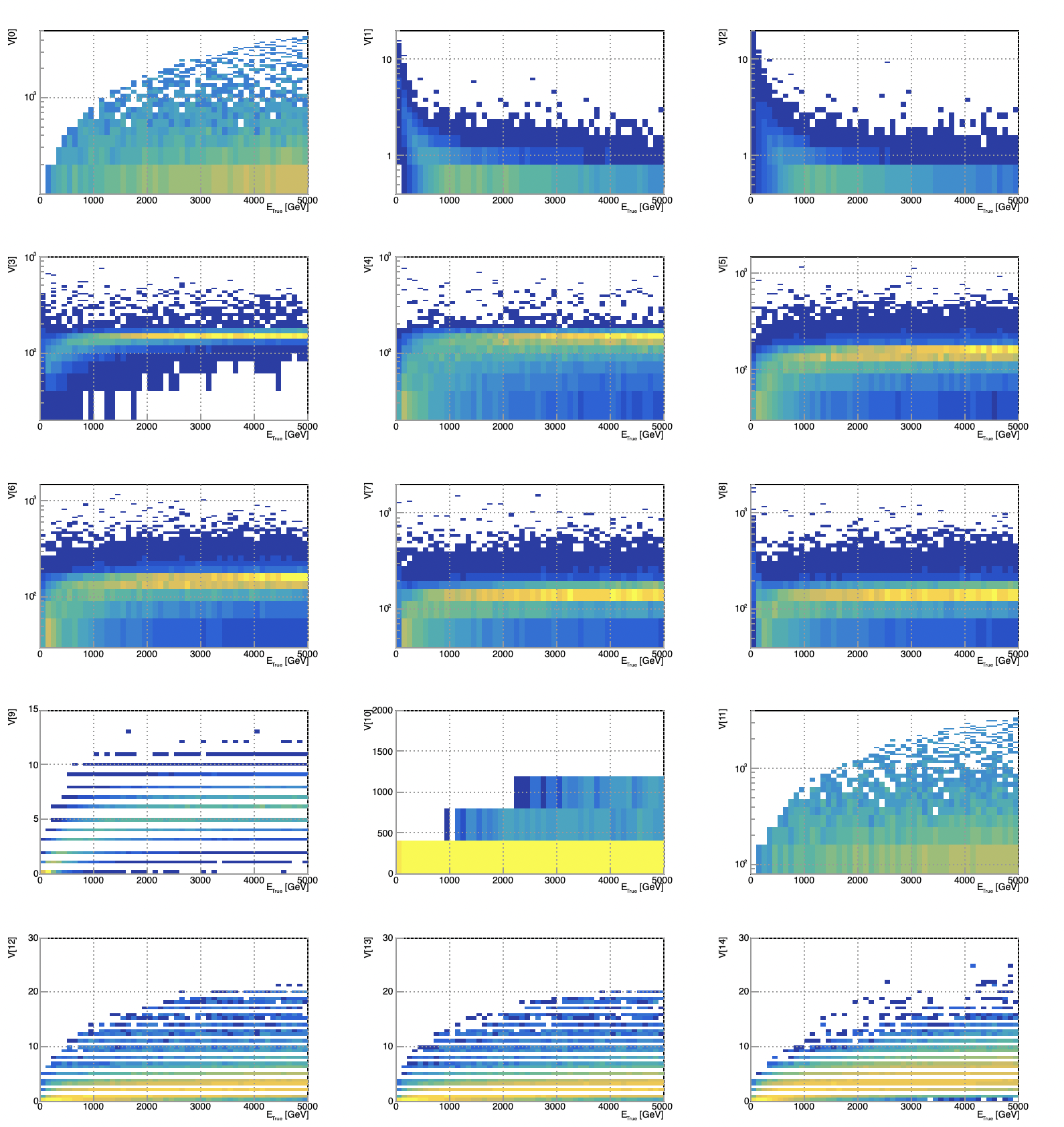}
     \caption{\em 2D histograms showing the dependence of the event features (on the y axes) on true muon energy (on the x axes). Top row: V[0],V[1], V[2]; second row: V[3],V[4], V[5]; third row: V[6],V[7],V[8]; fourth row: V[9],V[10],V[11]; bottom row: V[12],V[13],V[14]. Features are defined in Section~\ref{s:features}.}
        \label{f:feats}
    \end{center}
\end{figure}

\begin{figure}[ht]
 	\begin{center}
		  \includegraphics[width=\textwidth]{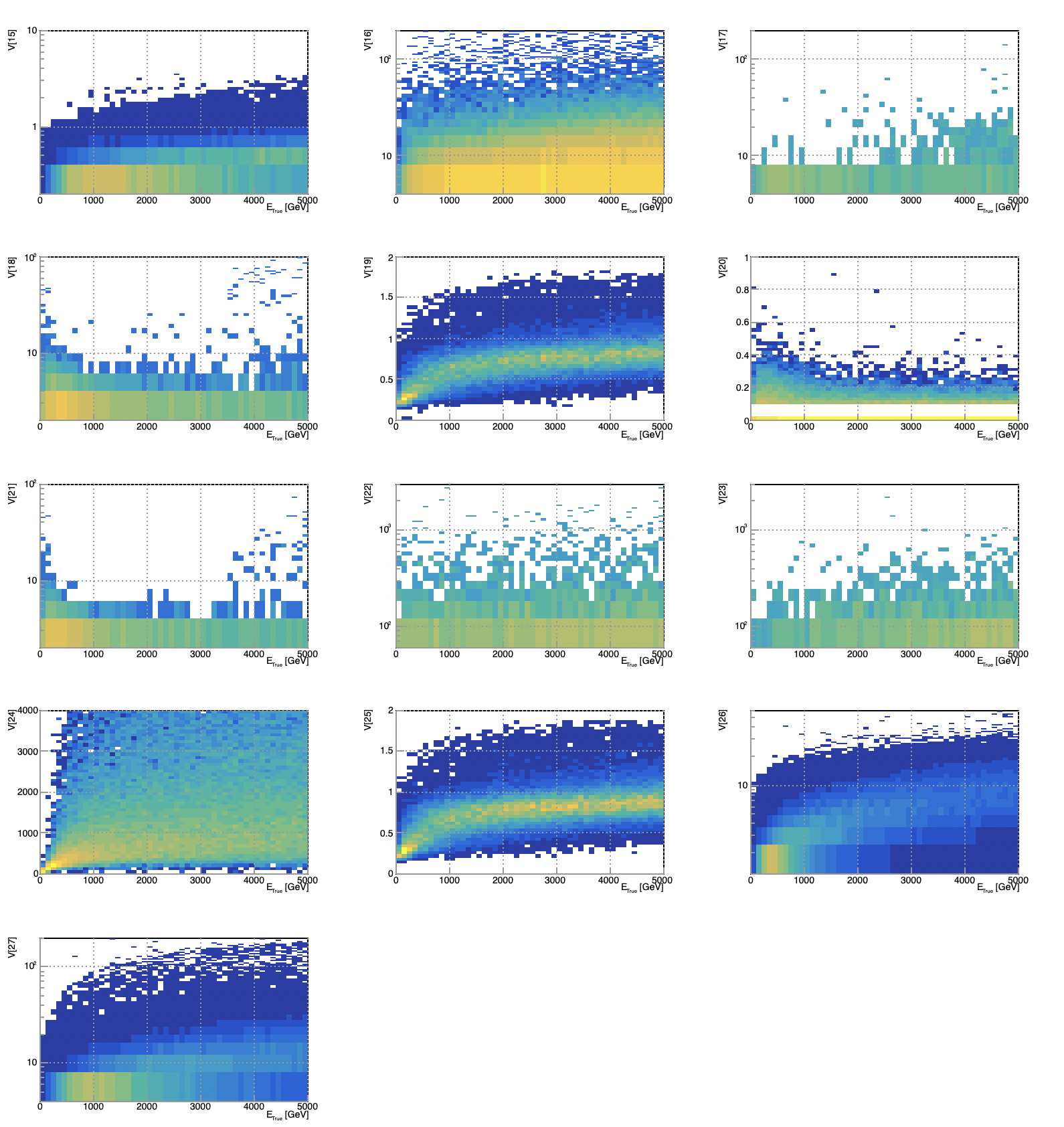}
     \caption{\em Marginals of event features and true muon energy. Top row: V[15],V[16], V[17]; second row: V[18],V[19], V[20]; third row: V[21],V[22],V[23]; fourth row: V[24],V[25],V[26]; bottom row: V[27]. Features are defined in Section~\ref{s:features}.}
        \label{f:feats}
    \end{center}
\end{figure}

\clearpage

\section{Variable pruning} \label{s:preprocessing} \label{s:pruning}

The original set of high-level features described in the previous section was designed to be fed into a neural network in order to provide an extensive and redundant description of observable characteristics (spatial distribution and intensity) of the energy deposits. As such, the whole set is unsuitable for use with a nearest-neighbor algorithm, which is sensitive to the curse of dimensionality. While a partial solution to the large dimensionality of the feature space is provided by random sampling of subspaces, as described in Sec.~\ref{s:knn}, we consider it beneficial to study the possibility of reducing the pool of informative variables, discarding the ones which are expected to contribute non-appreciably to the quality of a kNN-based regressor, or {\em a fortiori } those which --due to the larger dimensionality of the space when including them-- even worsen its performance.



The task of identifying what features to discard is however difficult. The complexity of the interdependencies of the 28-dimensional feature space makes the use of ordinary methods dubious: individual pairwise correlation coefficients, {\em e.g.}, will inform on the global behaviour of two variables, and thus often miss the presence of restricted regions of the variable spectrum where dependences are strong or absent. To exemplify this, we may mention the variable constructed on curvature information, $V[24]$ (see {\em supra}, Sec.~\ref{s:features}). That variable is quite informative of muon energy when muon energy is very small, because in that case energy deposits offer meaningful curvature information on the muon trajectory; on the other hand, at high muon energy the variable has absolutely zero information content. The overall correlation between that variable and muon energy is only very mild, yet the variable may still be quite useful in a kNN regression, as its value is very informative for a subset of the data. We show some results of correlation studies in the Appendix.

After studying several methods to assess the value of different variables we were most convinced by the study of the effect of individual variables on the regression results. We perform this test by running a large number of different regressions (1639), each of which employs a randomly chosen weak learner that investigates a different subspace of the feature space; we employ subspaces of dimension varying at random between 8 and 15 in these tests. From the regression results, for each variable we can then obtain the average value of the difference between the figure of merit (FoM) achieved by learners that use the variable and the FoM of learners that do not. Each regression employs a relatively small number (10,000) of training events and 50,000 events for testing, taken from the same training data sample of 700,000. No optimization of the weights and biases of training events is performed, to reduce CPU consumption of these tests; the rationale of this choice is that the relative importance of features should not be heavily coupled to the improvement in performance of the regression by the overparametrization provided by weights and biases of training data.

We rank each variable by its effect on each of the following four figures of merit. The first two, which we aim to minimize, are: the global loss per event, as defined in Sec.~ \ref{s:loss}, and the maximum resolution in regressed energy obtained after combining the regression result with a hypothetical independent tracking-based measurement with a $20\%$ relative uncertainty\footnote{This figure of merit and the ones that follow will be discussed in more detail in Sec.~\ref{s:regressor}}. The last two, which we aim to maximize, are: the discrimination power of the predicted energy, in number of standard deviations, between 2-TeV and 4-TeV muons, and the discrimination power of the predicted energy between 1-TeV and 3-TeV muons.  

Among the set of 28, we identify the six variables which rank the worst on each of the four figures of merit, by ordering them according to the difference between the FoM value achieved by learners that employ that variable and learners that do not. For loss and maximum resolution, this difference is negative (positive) if the variable has a positive (negative) effect in the regression; for the discrimination 2-4 TeV and 1-3 TeV a more negative difference indicates worse effect in the regression. The selection procedure results in a list of 11 variables, as many of them rank as the worst according to more than one criterion. Variables 4, 5, 6, 7, 3, 25 are the six worst ones based on the difference in loss; variables 7, 6, 5, 3, 8, 11 are the six worst ones based on the difference in max resolution; variables 7, 25, 6, 5, 18, 10 are the six worst ones based on the difference in discrimination 2-4 TeV; and variables 4, 20, 5, 7, 6, 11 are the six worst ones based on the difference in discrimination 1-3 TeV.

\begin{table}[h!]
\begin{center}
$\begin{array}{ c | c c c c }
\text{Variable} & \text{$\Delta$ Loss} & \text{$\Delta$ Max res. ($\times 10^4$)} & \text{$\Delta$ 2-4 TeV discr.} & \text{$\Delta$ 1-3 TeV discr.} \\
  \hline
  4 & 0.2939 \pm 0.0336 & \mathbf{-2.60 \pm 1.80} & -0.00105 \pm 0.00138 & -0.01427 \pm 0.00200 \\ 
  5 & 0.2531 \pm 0.0352 & 6.91 \pm 1.76 & -0.00432 \pm 0.00123 & -0.01120 \pm 0.00219 \\ 
  6 & 0.2299 \pm 0.0377 & 11.91 \pm 1.73 & -0.00505 \pm 0.00125 & -0.00803 \pm 0.00222 \\ 
  7 & 0.2034 \pm 0.0366 & 17.80 \pm 1.77 & -0.00770 \pm 0.00126 & -0.01002 \pm 0.00219 \\ 
  3 & 0.1646 \pm 0.0332 & 4.40 \pm 1.82 & -0.00079 \pm 0.00140 & -0.00357 \pm 0.00207 \\ 
  25 & 0.1575 \pm 0.0363 & 0.67\pm 1.86 & -0.00506 \pm 0.00139 & -0.00585 \pm 0.00215 \\ 
  8 & \mathbf{-0.0006 \pm 0.0365} & 3.40 \pm 1.81 & \mathbf{0.00303 \pm 0.00145} & \mathbf{0.01068 \pm 0.00221} \\ 
  11 & 0.0910 \pm 0.0341 & 2.93 \pm 1.74 & -0.00366 \pm 0.00116 & -0.00650 \pm 0.00197 \\ 
  18 & 0.0898 \pm 0.0346 & \mathbf{-1.02 \pm 1.69} & -0.00422 \pm 0.00116 & -0.00633 \pm 0.00198 \\ 
  10 & 0.0985 \pm 0.0331 & 2.61 \pm 1.73 & -0.00368 \pm 0.00118 & -0.00507 \pm 0.00193 \\ 
  20 & 0.0997 \pm 0.0359 &\mathbf{-2.91 \pm 1.764} & -0.00309 \pm 0.00116 & -0.01228 \pm 0.00203 
\end{array}$
\end{center}
\caption{The 11 variables identified as potentially non-useful by inspection of the effect of their inclusion or exclusion on the figures of merit. Rows contain, for each variable, the difference of the figures of merit (with their uncertainty) in learners that use the variable and learners that do not. Values highlighted in boldface show results which do not indicate a bad performance of the variable.}
\label{table:pruning2}
\end{table}

\noindent
From the results shown in Table~ \ref{table:pruning2} above, the variables with the worst impact in the regression performance are 4, 5, 6, 7, 3, 25, 8, 11, 18, 10, 20. We can also observe in the table that the inclusion of variable 8 in weak learners does not improve maximum resolution, but improves all other figures of merit. Because of this, we keep it in the variable set; all other variables are instead removed. The final set of 18 variables on which the algorithm is run is therefore: 0, 1, 2, 8, 9, 12, 13, 14, 15, 16, 17, 19, 21, 22, 23, 24, 26, 27. 

It is also worth noting that the test produces overall coherent results as the identified variables mostly worsen all the figures of merit (by having positive values of loss and maximum resolution difference, and negative values of the discrimination powers). To provide a point of comparison, we list below the differences in values obtained by the best performing variables in the same test, selecting the three which perform best in each of the four figures of merit: 9, 1, 2 according to the difference in loss; 9,26,1 according to the maximum resolution; 26, 9, 14 according to the 2-4 TeV discrimination; and 9, 2, 1 according to the 1-3 TeV discrimination. This result in five variables, as shown in Table~\ref{table:pruning3}.

\begin{table}[h!]
\begin{center}
$\begin{array}{ c | c c c c }
\text{Variable} & \text{$\Delta$ Loss} & \text{$\Delta$ Max res. ($*10^4$)} & \text{$\Delta$ 2-4 TeV discr.} & \text{$\Delta$ 1-3 TeV discr.} \\
  \hline
9  & -0.977 \pm 0.027 & -23.56 \pm 1.53          & 0.0166 \pm 0.0011 & 0.0528 \pm 0.0017 \\
1  & -0.572 \pm 0.037 &  -4.66 \pm 1.78          & 0.0076 \pm 0.0014 & 0.0260 \pm 0.0022 \\
2  & -0.492 \pm 0.039 &  -1.03 \pm 1.80          & 0.0083 \pm 0.0012 & 0.0347 \pm 0.0021 \\
26 & -0.135 \pm 0.036 & -22.94 \pm 1.81          & 0.0172 \pm 0.0015 & 0.0127 \pm 0.0022 \\
14 & -0.208 \pm 0.033 &   \mathbf{1.16 \pm 1.74} & 0.0135 \pm 0.0012 & 0.0044 \pm 0.0019 \\
\end{array}$
\end{center}
\caption{The 5 variables identified as most useful by inspection of the effect of their inclusion or exclusion on the figures of merit. Rows contain, for each variable, the difference of the figures of merit (with their uncertainty) in learners that use the variable and learners that do not. Values highlighted in boldface show results which do not indicate a good performance of the variable.}
\label{table:pruning3}
\end{table}



To validate the results shown {\em supra} we run a regression using a single learner, on a training dataset independent from the one used in the ones used above. The regression is first performed using all the 28 variables, using 10,000 training and 50,000 test events. Then the regression is performed with identical settings by removing some of the variables previously identified as non-informative.

\begin{table}[h!]
\begin{center}
\begin{tabular}{ c | c c c c }
 Excluded variables & Loss & Max res. & 2-4 TeV discr. & 1-3 TeV discr. \\ 
\hline
 None & 38.4141 & 0.330749 & 0.706453 & 1.12254 \\ 
 3, 4, 5, 6 & 37.5701 & 0.32778 & 0.709819 & 1.15602 \\  
 3, 4, 5, 6, 7, 25 & 37.2649 & 0.327508 & 0.715047 & 1.1493 \\ 
 3, 4, 5, 6, 7, 25, 10, 18, 20 & 37.3583 & 0.325874 & 0.718947 & 1.15443    
\end{tabular}
\end{center}
\caption{Values of the figures of merit resulting from regressions employing gradually reduced variable sets. See the text for details.}
\label{table:pruning1}
\end{table}

\noindent
The figures of merit in Table~\ref{table:pruning1} show that the removal of non-informative variables improves the performance of the algorithm. However, the large dimensionality of the explored spaces (18 to 28 dimensions) might suggest that the results are due to the positive effect of reducing the dimensionality of the scanned subspaces. We therefore perform one further test, where we study lower-dimensional subspaces spanned by variables 9, 1, 2, 14, 26 (the best performing ones) and adding one at a time the variables we identified as non-informative. In these regressions we employ 10,000 training and 50,000 testing events. The results of this further test are shown in Table~ \ref{table:pruning4}. We observe a gradual degradation of the loss (from 37.40 when not using any of the 10 pruned variables, to 38.97 when using all of them), and a gradual decrease of the 1-3 TeV discrimination (from 1.20 to 1.14 standard deviations). The other figures of merit show less significant trends. Overall, this further test confirms the advantage of neglecting the 10 pruned variables.

\begin{table}[h!]
\begin{center}
\begin{tabular}{ c | c c c c }
Added variables & Loss & Max res. & 2-4 TeV discr. & 1-3 TeV discr. \\ 
\hline
 0  & 37.40   & 0.3306 & 0.6486 & 1.1972 \\
 1  & 37.77   & 0.3282 & 0.6458 & 1.1787 \\  
 2  & 38.06   & 0.3276 & 0.6580 & 1.1714 \\
 3  & 38.31   & 0.3272 & 0.6583 & 1.1674 \\
 4  & 38.46   & 0.3276 & 0.6536 & 1.1567 \\
 5  & 38.69   & 0.3283 & 0.6459 & 1.1438 \\
 6  & 38.73   & 0.3270 & 0.6498 & 1.1444 \\
 7  & 38.70   & 0.3264 & 0.6526 & 1.1451 \\
 8  & 38.70   & 0.3266 & 0.6518 & 1.1451 \\
 9  & 38.85   & 0.3280 & 0.6503 & 1.1449 \\
 10 & 38.97   & 0.3285 & 0.6465 & 1.1384 \\
\end{tabular}
\end{center}
\caption{Values of the figures of merit resulting from regressions employing gradually increased variable sets starting from the one constituted by the most performing variables (9, 1, 2, 14, 26) and adding one non-informative variable at a time (in the order 4, 5, 6, 7, 3, 25, 11, 18, 10, 20). See the text for other details.}
\label{table:pruning4}
\end{table}

\clearpage
\section{The regressor}\label{s:regressor}

As discussed in Sec.~\ref{s:knn}, the kNN regressor we developed uses feature sub-sampling ~\cite{featsample} as a source of stochasticity to build an ensemble of weak learners, whose relative importance (parametrized by $N_{wl}$ weights $W_{wl}$ adding up to 1) is learnt via gradient descent. The largest part of the learners flexibility is however obtained by associating to each training event, and for each weak learner, a bias and a weight. By optimizing via gradient descent these parameters, the system becomes highly overparametrized. In the following section the specific details of the algorithm relevant for the muon energy regression task are described in detail.

\subsection{The algorithm}\label{s:algorithm}

The algorithm employs 18 variables selected as discussed in Sec.~\ref{s:pruning} from the original 28 described in Sec.~\ref{s:features}. An optimization of the number of neighbors indicates that $k=100$ is a reasonable choice for the regression. Values above 100 correspond to significant increases in processing time as well as in non-local averages of muon energies, given the relatively small number of training events (300,000-400,000 events are used for the final results) and large dimensionality of the investigated subspaces (typically 8-12 dimensions); on the other hand, we observed that if $k$ becomes smaller than about 50 the precision of the estimate starts to degrade, as is common for kNN regressors. A choice of $k=100$ with respect to {\em e.g.} $k=50$ has in the case at hand the benefit that more events concur in the prediction of each test event, thus contributing with a larger number of parameters (typically a bias and a weight for each event and each of 5-20 weak learners, for a total of a few thousand parameters per test point).

\subsubsection{Weak learners definition}

The regression task starts with a definition of the sets of weak learners employed in a run. This number linearly affects the computing time, which for a run of 400,000 training events and 100,000 test events may correspond to several days of running. We chose to identify a total of 32 sets of weak learners: 8 sets of five learners, 8 sets of 10 learners, 8 sets of 15 learners, and 8 sets of 20 learners. Each set was separately determined by optimization searches running on cross-validation sets of training data, without an optimization of weights and biases of training events. The procedure for the definition of the weak learners of each set is as follows:

\begin{itemize}
    \item A bootstrapped set of 10,000 training events is selected
    \item $N_{wl}$ weak learners are defined at random, by selecting a variable fraction
    (from 30\% to 80\%) of the active space dimensions among the 18 available
    \item An iterative loop is executed 100 times. At each step, a small fraction of the flags defining the subspace of each weak learner is changed at random; then the regression is performed on a batch of 5000 events belonging to the remainder of the overall training sample (700,000 events) from the ones employed for training; the loss is computed and compared to previous results. If it is smaller, the new set of weak learners substitutes the former one.
\end{itemize}

\noindent
The resulting set of weak learners collectively perform significantly better than the original random set; typically the loss function decreases by 2 to $5\%$ with the procedure. We have experimented at length with less stochastic methods for learning better weak learners, such as genetic breeding and other techniques, but we found no advantage in these methods on our specific problem. 

Table~ \ref{t:NL5} shows as an example the flags identifying a combination of five weak learners obtained by the procedure described above.

\begin{table}[h!]
\begin{center}
\begin{tabular}{lc}
Flags of weak learner in pool (0:27) & Individual loss of WL\\
\hline
1 1 1 0 0 0 0 0 0 1 0 0 0 0 1 0 1 0 0 1 0 0 0 1 1 0 1 0 &37.22\\
0 1 1 0 0 0 0 0 1 1 0 0 0 1 1 1 0 0 0 0 0 1 0 0 1 0 1 1 &37.37\\
1 1 1 0 0 0 0 0 0 1 0 0 1 0 1 0 1 0 0 0 0 1 0 0 0 0 1 1 &37.34\\
1 0 1 0 0 0 0 0 0 1 0 0 1 0 0 0 0 0 0 0 0 0 1 1 1 0 1 1 &37.60\\
1 1 1 0 0 0 0 0 0 1 0 0 0 0 1 0 0 0 0 0 0 0 0 1 0 0 0 0 &37.53\\
\hline
Global loss for this batch  & 37.02 \\
\end{tabular}

\caption{Active subspace dimensions for five weak learners in a 5-WL set.}
\label{t:NL5}
\end{center}
\end{table}

\subsubsection{Hyperparameters}

In addition to a specification of the subspaces where the nearest neighbors are found for each test point, the regressor necessitates of a number of hyperparameters that define its behavior. We list them below, including for completeness ones we already mentioned {\em supra}. 

\begin{itemize}
    \item $k$, the numbers of neighbors used by the algorithm to compute the average energy of training events;
    \item a learning rate is also defined to decide the width of the gradient descent steps in the value of weights and biases; this is done by defining the value of two multipliers $\lambda_w$, $\lambda_b$ of the partial derivative of the loss with respect to training event weights and biases. The optimization of these values is a non-trivial, separate task which was performed independently. 
    \item the learning rate is increased during minimization whenever the direction of steepest descent remains the same, and is reset to an exponentially decreasing value when the maximum gradient changes direction.
    \item $N_{batch}$ is the number of events considered during the optimization cycle.
    \item $\alpha_0$, $\alpha_1$ determine the relative strength of the MSE-like term and the nonlinear-response-penalization term (see Sec.~2).
\end{itemize}

\noindent
Other parameters are described below. Our processing of the available data for the regression task proceeds as follows:\par

\begin{enumerate}
    \item Input data are standardized by subtracting means $\hat{V_n}$ and dividing each of the $n=1, ... , 18$ features remaining after the pruning described in Sec.~\ref{s:pruning} by their all-sample root-mean-square, $RMS_n$, of the corresponding marginal distribution, so that $V_n=(V_n-\hat{V_n})/RMS_n$.
    \item Data are divided into prediction, training, and test samples. The test sample is kept separated, and it amounts to 187,000 events in the $[0.05~\mathrm{TeV}, 8~\mathrm{TeV}]$ range --hence only about $60\%$ of it can be used for tests in the $[0.05~\mathrm{TeV}, 5~\mathrm{TeV}]$ region of interest for our task. Training and optimization sets are only defined at run time and depend on the kind of cross-validation scheme; they may contain, {\em e.g.}, up to $(N_{\text{train}}, N_{\text{opt}}) =$ (400,000, 300,000) events. 
    \item An ensemble of $N_{wl}$ weak learners is created through feature sub-sampling by randomly setting to 0 or 1 the indicator function $I_{wl}(i)$.
    So, {\em e.g.}, if an indicator function is $I_{wl}=(1, 1, 1, 1, 0, ... , 0)$, the corresponding \kNN regressor works in the 4-dimensional subspace spanned by the first four variables. The number of learners $N_{wl}$ is chosen at run time, and so is the average number $N_{active}$ of dimensions set to 1 for each learner.
    \item An optimization cycle is run using training data ($j=1, ..., N_{train})$, to find the best value of weights $W_i$ ($i=1, ..., N_{wl}$) such that a linear combination of the predictions $P_{j,i}$ of the $N_{wl}$ learners, $P_j=\sum{W_i P_{j,i}}$ produces the lowest value of a loss function (the definition of the loss and the details of the optimization cycle are given {\em infra}); the $N_{wl}$ weights are constrained to add up to one, but are independently allowed to be negative, adding flexibility to the estimators.
    \item Weights and biases, along with optimal weak learners and their weights, can finally be used to run on an independent test sample.
\end{enumerate}

\subsubsection{Weights and biases optimization}

Weights for every training event are initialized to a value which depends on the true energy of the event, to downscale the impact of high-energy events (which in the training sample range up to 8 TeV, hence considerably higher than the range of events ($[0.05~\mathrm{TeV}, 5~\mathrm{TeV}]$) which contribute to the loss evaluation) according to a sigmoid function, exactly as is done in the analysis performed in ~\cite{cnnpaper}: we set 

\begin{equation}
x_i = \frac{E_i-5}{0.3(1+Ind(E_i>5))}
\end{equation}

\noindent
where $E_i$ is the true muon energy of event $i$ in TeV, and $Ind(E_i>5)$ equals 1 if true muon energy is above 5 TeV, and 0 otherwise. We then get, for all weak learners $l=1...N_{wl}$ and event $i$, \par

\begin{equation}
    w(l,i) = 1-\frac{1}{1+e^{-x_i}}.
\end{equation}

\noindent
Weights as defined {\em supra} effectively act as a prior density on the muon energy, making high-energy predictions above 5 TeV less probable. In fact, in any realistic application one such prior would be useful -{\em e.g.}, in a 4-TeV electron-positron collider one would rather set this as a step function sharply ending at 4 TeV, or even lower; the smooth function we chose in ~\cite{cnnpaper} and adopt here is meant to influence very little the regression task, while avoiding a too large impact on the chosen cutoff in the generated muon energy range.

Biases are instead initially set to zero for all training event and all weak learners:

\begin{equation}
    b(l,i) = 0.
\end{equation}

\noindent
To optimize weights and biases of each training event, separately for each weak learner in a set, we divide the 700,000 training events set into a set of 400,000 which we use to train the and  regressors with $N_{wl}$ = 5 and the ones with $N_{wl}$ = 10, and a set of 300,000 which we use to train the regressors with $N_{wl}$ = 15 and the ones with $N_{wl}$ = 20. In each case, the remaining events are used for 36 loops of gradient descent, for each of the 32 regressors. Within each loop, we select batches of 5000 events which we use to compute derivatives of the loss with respect to the weights and biases of each training event, as well as relative weights of the $N_{wl}$ learners in a regressor. As these parameters are updated we monitor the two components of the loss function discussed in Sec.~\ref{s:loss}, which are observed to smoothly decrease in all cases. We have verified, by monitoring the loss on independent samples during training, that the number of epochs of this optimization cycles is well clear of overtraining --in fact, there are indications that the regressors are undertrained. The limitation comes from CPU availability.

\begin{figure}[h!]
\begin{center}
\includegraphics[width=15cm]{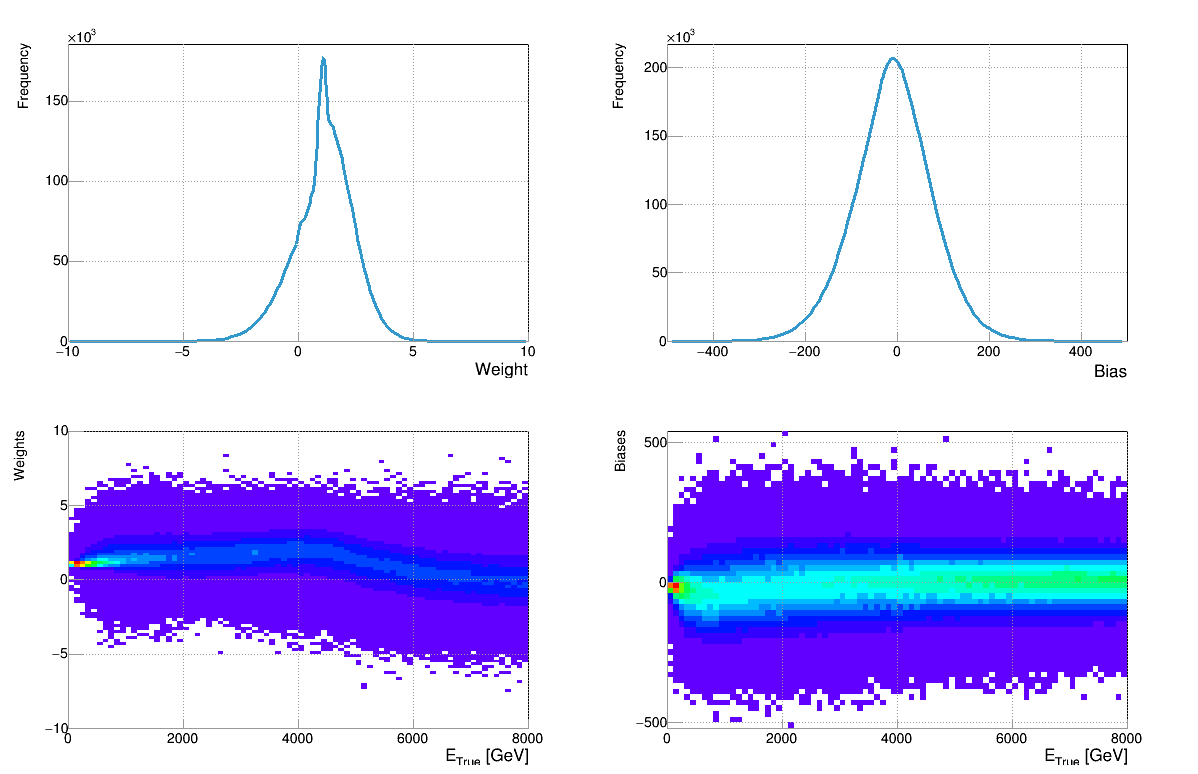}
\caption{Example of the distribution of weights obtained by an optimization run with 400,000 training events and 5 learners. Top left: distribution of event weights; top right: distribution of event biases (in GeV). Bottom left: weights versus true muon energy; bottom right: biases (in GeV) versus true muon energy.}
\label{f:weightsandbiases}
\end{center}
\end{figure}

After completion of the above procedure, we may study the distribution of training event weights and biases, and their relation with the true muon energy. An example of these distributions are shown in Fig.~\ref{f:weightsandbiases}.

\subsubsection {Final run and averaging}

At the end of the optimization procedure for each of the 32 regressors, we choose two performers from each set of eight for our final test runs. The choice is loosely based on the four figures of merit already described in Sec.~\ref{s:pruning}, when we give preference to high discrimination power between muon energies. The figures of merit resulting from the 32 optimization runs are shown in Table~\ref{t:32runs}.

\begin{table}[h!]
\begin{center}
\begin{tabular}{l c c c c c}
ID & WL & Loss & MaxRes & Discr24 & Discr13 \\
\hline
1001 & 5  & 34.96 & 0.3487 & 0.6668 & 1.2799  \\
1002 & 5  & 34.79 & 0.3434 & 0.6574 & 1.2756  \\
1003 & 5  & 35.07 & 0.3490 & 0.6571 & 1.2551  \\
\bf{1004} & 5  & 35.02 & 0.3447 & 0.7004 & 1.3048  \\
1005 & 5  & 34.87 & 0.3458 & 0.6699 & 1.2870  \\
1006 & 5  & 34.90 & 0.3450 & 0.6634 & 1.2794  \\
\bf{1007} & 5  & 34.88 & 0.3383 & 0.6893 & 1.2854  \\
1008 & 5  & 34.95 & 0.3475 & 0.6610 & 1.2915  \\
\bf{1009} & 10 & 34.40 & 0.3479 & 0.6859 & 1.3200 \\
1010 & 10 & 34.78 & 0.3450 & 0.6734 & 1.3040 \\
1011 & 10 & 34.53 & 0.3418 & 0.6856 & 1.3112 \\
1012 & 10 & 34.66 & 0.3393 & 0.6803 & 1.3202 \\
\bf{1013} & 10 & 34.55 & 0.3446 & 0.6861 & 1.3408 \\
1014 & 10 & 34.53 & 0.3410 & 0.6743 & 1.3067 \\
1015 & 10 & 34.58 & 0.3415 & 0.6828 & 1.3066 \\
1016 & 10 & 34.50 & 0.3423 & 0.6849 & 1.3044 \\
1017 & 15 & 34.23 & 0.3402 & 0.7069 & 1.3206 \\
\bf{1018} & 15 & 34.15 & 0.3395 & 0.7241 & 1.3421 \\
\bf{1019} & 15 & 34.12 & 0.3427 & 0.7156 & 1.3326 \\
1020 & 15 & 34.19 & 0.3380 &  0.7077 & 1.3354 \\
1021 & 15 & 34.26 & 0.3429 &  0.6921 & 1.3328 \\
1022 & 15 & 34.32 & 0.3420 & 0.7063 & 1.3225 \\
1023 & 15 & 34.23 & 0.3452 & 0.7086 & 1.3255 \\
1024 & 15 & 34.20 & 0.3397 & 0.7077 & 1.3296 \\
1025 & 20 & 34.20 & 0.3413 & 0.6942 & 1.3454 \\
1026 & 20 & 34.19 & 0.3374 & 0.7088 & 1.3403 \\
1027 & 20 & 34.39 & 0.3426 & 0.6883 & 1.3295 \\
1028 & 20 & 34.15 & 0.3423 & 0.6962 & 1.3329 \\
\bf{1029} & 20 & 34.10 & 0.3374 & 0.7231 & 1.3418 \\
1030 & 20 & 34.26 & 0.3451 & 0.6979 & 1.3520 \\
1031 & 20 & 34.15 & 0.3407 & 0.7021 & 1.3357 \\
\bf{1032} & 20 & 34.12 & 0.3413 & 0.7084 & 1.3496 \\
\end{tabular}
\end{center}
\caption {Figures of merit resulting from optimization runs on the 32 regressors. The ones chosen for the final testing run are in boldface. See the text for detail.}
\label{t:32runs}
\end{table}

\noindent
From the result of the optimization runs, we select the following sets of regressors: those with ID 1004 and 1007 with 5 weak learners, those with ID 1009 and 1013 with 10 weak learners, those with ID 1018 and 1019 with 15 weak learners, and those with ID 1029 and 1032 with 20 weak learners.

A final run with 80,000 test events 
is performed to obtain the prediction of each of the eight regressors. The four regressors with $N_{wl}$ = 5 and $N_{wl}$ = 10 base their prediction on 400,000 training events, together with the associated weights and biases; and the four with $N_{wl}$ = 15 and $N_{wl}$ = 20 employ the remainder 300,000 events of the training set, and associated weights and biases. The total number of parameters for these final runs is of $N_{weights} = (2 \times 5 + 2 \times 10) \times 4 \times 10^5 + (2 \times 15 + 2 \times 20) \times 3 \times 10^5 = 33M$, plus an equal number of bias values. The total number of parameters of the regression task in its entirety is therefore 66 million.

\subsection {Results}\label{s:tests}

In this section we show the final results of the regression described {\em supra}, and we compare them with those of the NN and CNN regressors described in ~\cite{cnnpaper}, as well as to those of a non-optimized kNN regressor and of a XGBoost decision tree.

\begin{figure}[h!]
\begin{center}
\includegraphics[width=\textwidth]{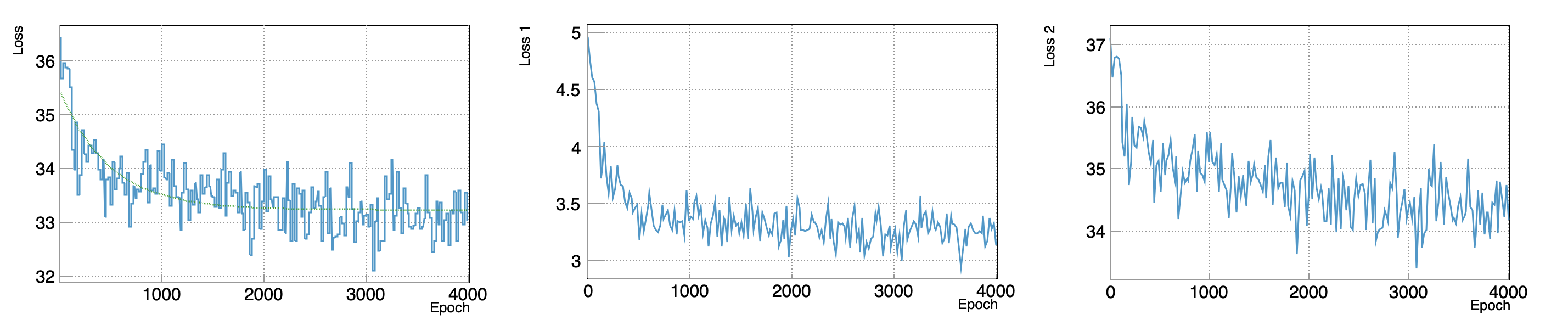}
\caption{Evolution of the loss function and its components with the gradient descent iterations (epochs). Left: total loss function ($L$); center: non-linearity penalization term ($L_2$), right:  modified MSE term ($L_1$). The two terms on the center and right do not add up to the one on the left because of a rescaling factor.}
\label{f:loss}
\end{center}
\end{figure}

The final loss of the averaged regressors, on 80,000 test events, results in $L=33.734$, which is smaller than any of the losses of the eight inputs. The other figures of merit are: $MaxRes = 0.3372$; $Max Area = 0.2154$; $Discr_{24}=0.7025$; $Discr_{13} = 1.3318$. Figure \ref{f:loss} shows a sample training of the regressor and the coherent decrease of the two components of the loss function.

\begin{figure}[h!]
\begin{center}
\includegraphics[width=\textwidth]{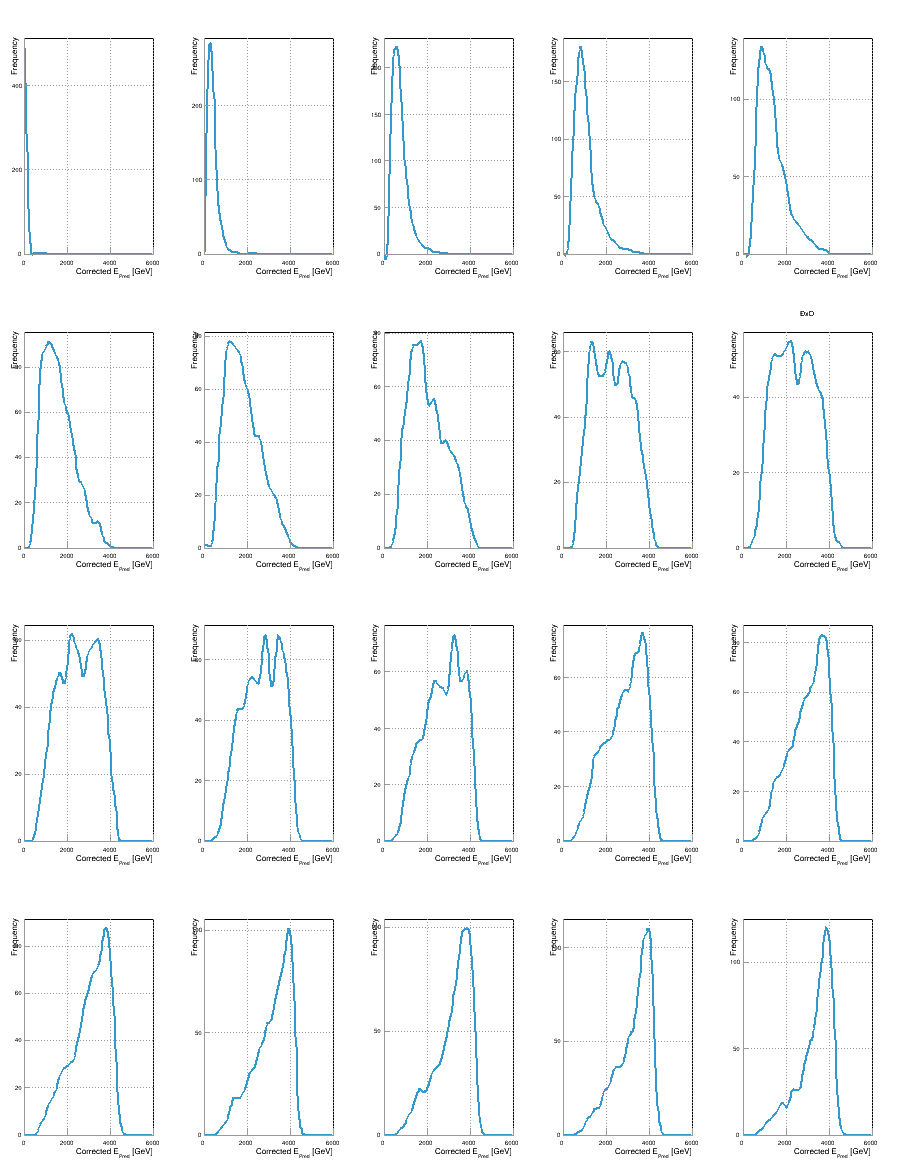}
\caption{Marginal distributions of predicted muon energy for different ranges of true muon energy. From top to bottom and left to right the true energy increases by 0.2 TeV from 0.1 to 3.9 TeV.}
\label{f:eprc}
\end{center}
\end{figure}

\begin{figure}[h!]
\begin{center}
\includegraphics[width=8cm]{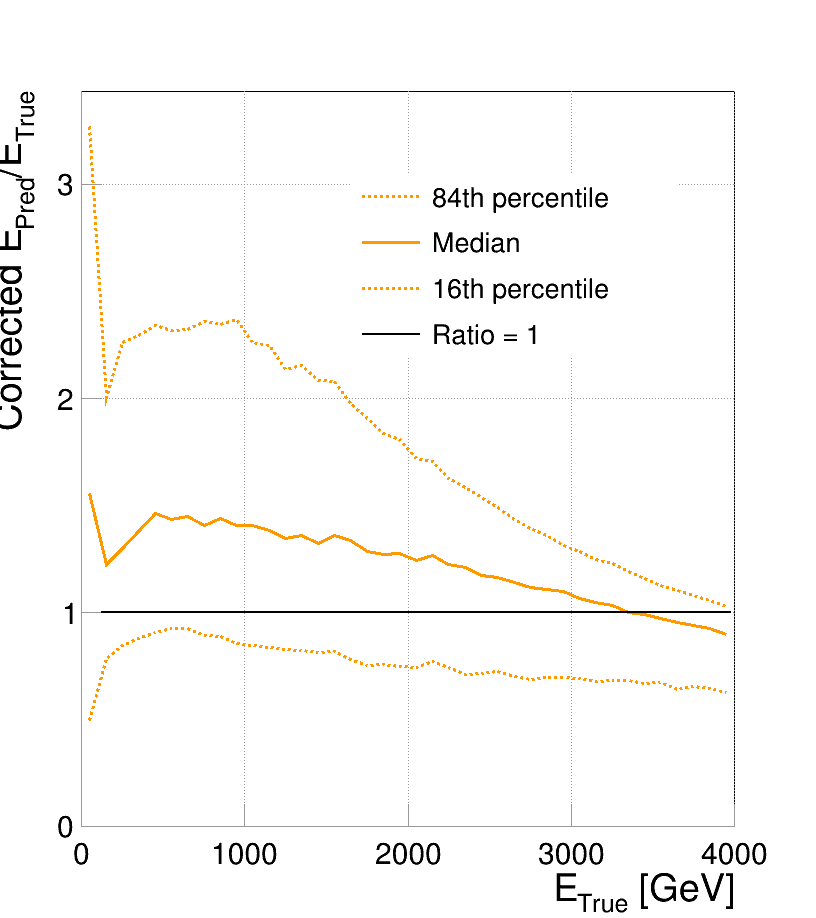}
\caption{Median, $84^{th}$, and $16^{th}$ percentiles of the ratio of corrected predicted energy and true energy versus true energy.}
\label{f:rr}
\end{center}
\end{figure}

In Fig.~\ref{f:eprc} we show marginal distributions of predicted energy for muon true energy varying from 0.1 to 3.9 TeV in 0.2 TeV intervals. One observes a marked non-Gaussianity of these distributions --a hint of the the complexity of the problem. Figure~\ref{f:rr} further shows that a bias in the prediction is still present, notwithstanding the penalization term in the likelihood; the bias is however smaller than the RMS and its contribution to the MSE is relatively small.

\begin{figure}[h!]
\begin{center}
\includegraphics[width=\textwidth]{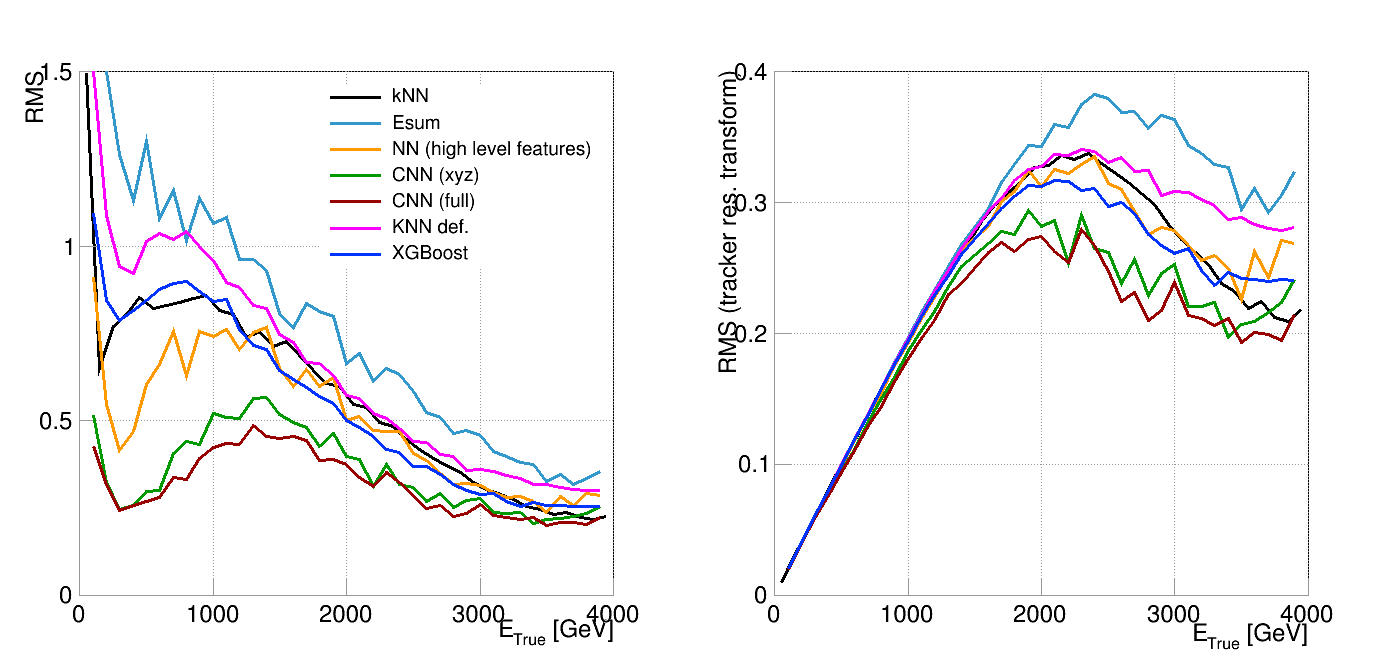}
\caption{Left: comparison of the mean squared error of predictions of different algorithms employing the same training and test data. Right: comparisons of the mean squared error of the combinations of tracker and calorimeter regressed predictions. Black: deep regression kNN (described in this article); light blue: energy sum model; orange: neural network with high level features; green: CNN with spatial features; red: full CNN; magenta: classical kNN; blue: XGBoost.}
\label{f:comp}
\end{center}
\end{figure}

As shown in Fig.~\ref{f:comp}, the deep kNN regressor we developed has performance comparable to those of XGBoost and of a multi-layer neural network running on the same input features and datasets, and significantly outperforms a regular kNN regressor. The convolutional neural network we developed in~\cite{cnnpaper}, on the other hand, still is in another league, as the MSE of its predictions is from $30\%$ to $50\%$ smaller than those of all other results. On the other hand, it needs to be pointed out that the CNN has access to full granularity information on the energy deposits in the calorimeter for each muon event --hence the comparison is only illustrative of the amount of information that can further be mined from that fine-grained distribution.

\begin{figure}[h!]
\begin{center}
\includegraphics[width=\textwidth]{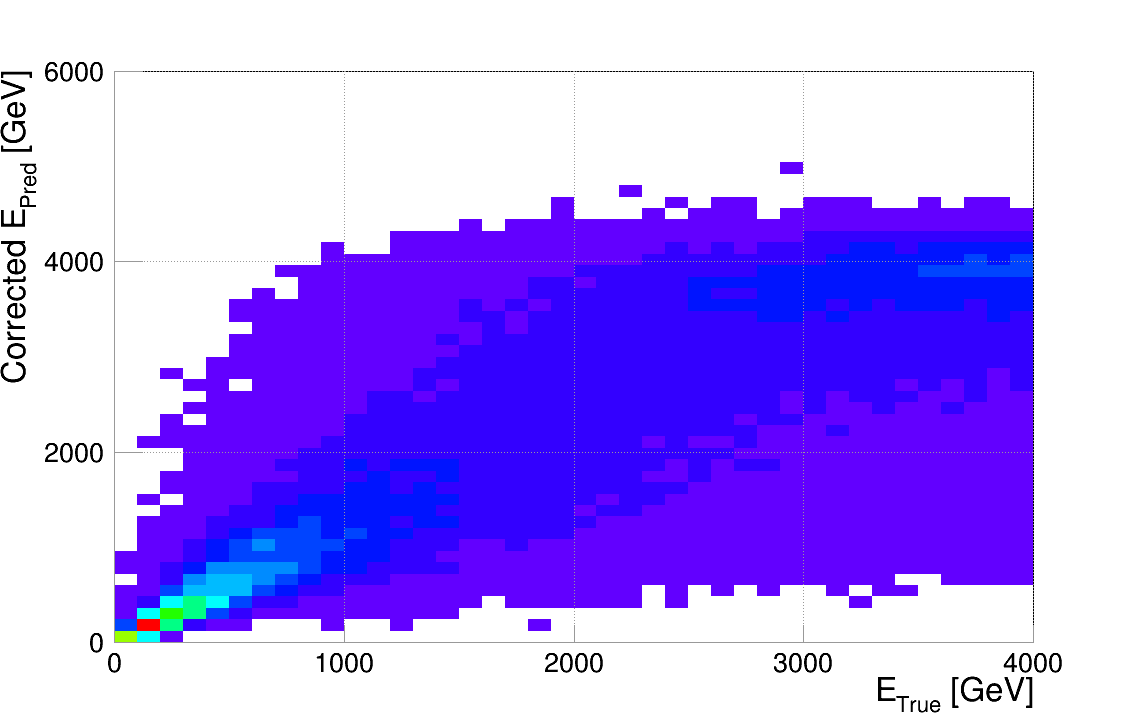}
\caption{Distribution of the corrected predicted muon energy versus true muon energy.}
\label{f:eec}
\end{center}
\end{figure}

\begin{figure}[h!]
\begin{center}
\includegraphics[width=\textwidth]{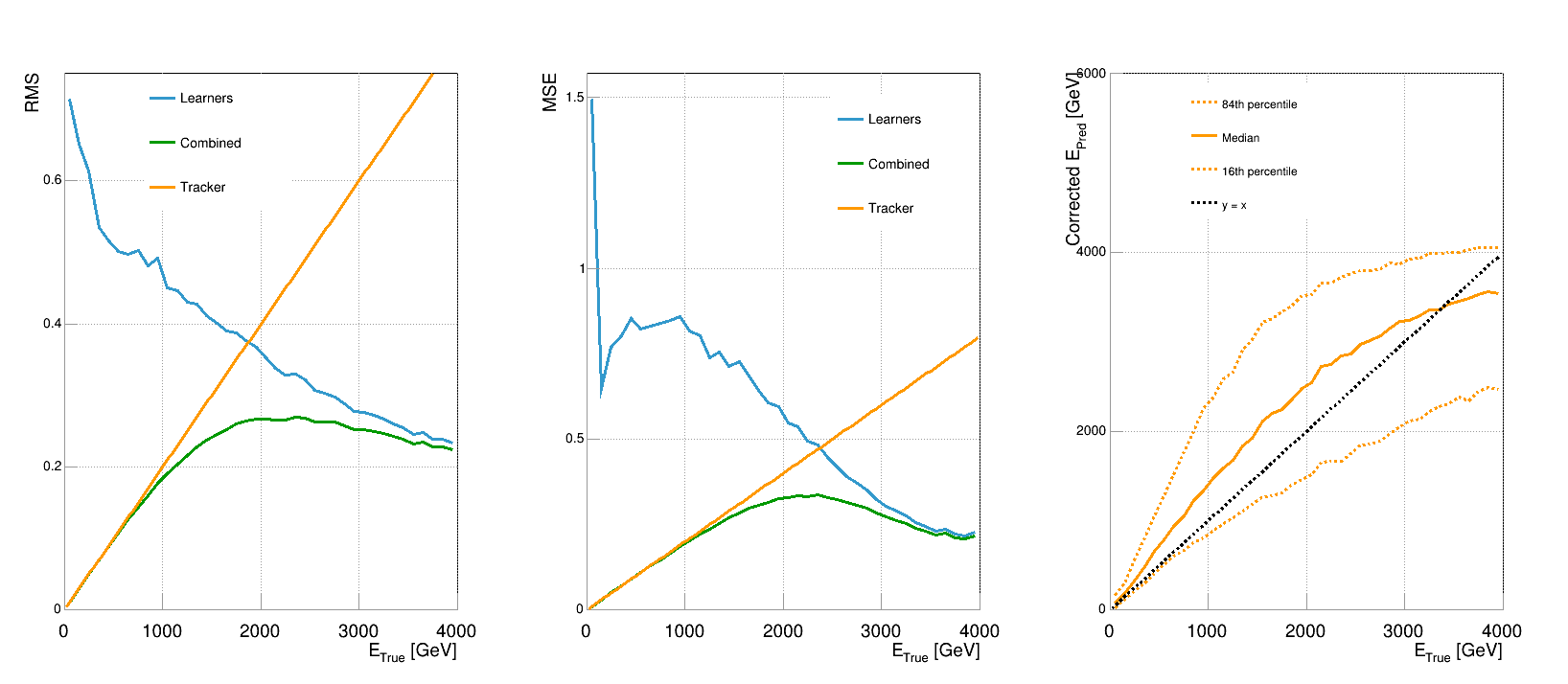}
\caption{Left: RMS resulting from learners-only, tracker-only and combined predictions, against true muon energy; center: MSE resulting from learners-only, tracker-only and combined predictions, against true muon energy; right: median, $84^{th}$, and $16^{th}$ percentiles of corrected predicted energy against true muon energy.}
\label{f:mm}
\end{center}
\end{figure}

\clearpage 
\section{Conclusions}\label{s:conclusion}

Moved by curiosity over the potential of the old-fashioned kNN statistical learning technique when compared to today's titanic deep learning architectures, in this work we have tried to determine how far can a kNN regression algorithm be pushed in its over-parametrization of a multi-dimensional problem, and how competitive it can be with today's state of the art algorithms. The occasion for doing so was the investigation of the problem of muon energy measurement in a future calorimeter for a high-energy machine, when curvature information may prove insufficient to provide the desirable resolution. In ~\cite{cnnpaper} we approached that question using a deep, customized convolutional neural network, and proved that $20\%$ relative resolution can be obtained for multi-TeV muon energy in a very fine grained calorimeter. In this paper we leverage those results to provide a point of comparison to what we can achieve by optimizing the large number of parameters of a customized nearest neighbor algorithm. Our results show that performance comparable to that of boosted decision trees and neural networks is achievable despite the complicated, highly multi-dimensional nature of the data. 

One caveat easily revealed by the practical implementation of our solution is the very large CPU load. Even omitting consideration of all preliminary studies of hyperparameters and other tests, the direct extraction of presented results requires several months of CPU on a single machine. It was only possible to compress that time to a few weeks by parallelizing the task, but it is clear that if CPU time is a metric the nearest neighbor technique cannot compare to the newer methods. On the other hand, it is to our knowledge the first time that a nearest neighbor can obtain competitive results by employing ${\cal{O}}(10^8)$ free parameters all individually optimized by stochastic gradient descent means. The interest of this observation is purely academic, but not negligible in our opinion.

\clearpage
\section*{Appendix 1: Correlation study}

In Sec.~\ref{s:pruning} we argued that linear correlation tests may not be suitable for the selection of non-informative variables for the kNN regression task. Here we show the results of those tests. We compute the correlation (Pearson and Spearman) between each variable and the true muon energy, the correlation (Pearson) between variables, the variance of the normalized variables; we identify variables that are very correlated with other variables, which might thus be argued to not offer new information; we also identify ones that are non correlated with true muon energy, as well as  ones that have very small variance and thus potentially contain little information.

\begin{figure}[ht]
 	\begin{center}
		  \includegraphics[width=\textwidth]{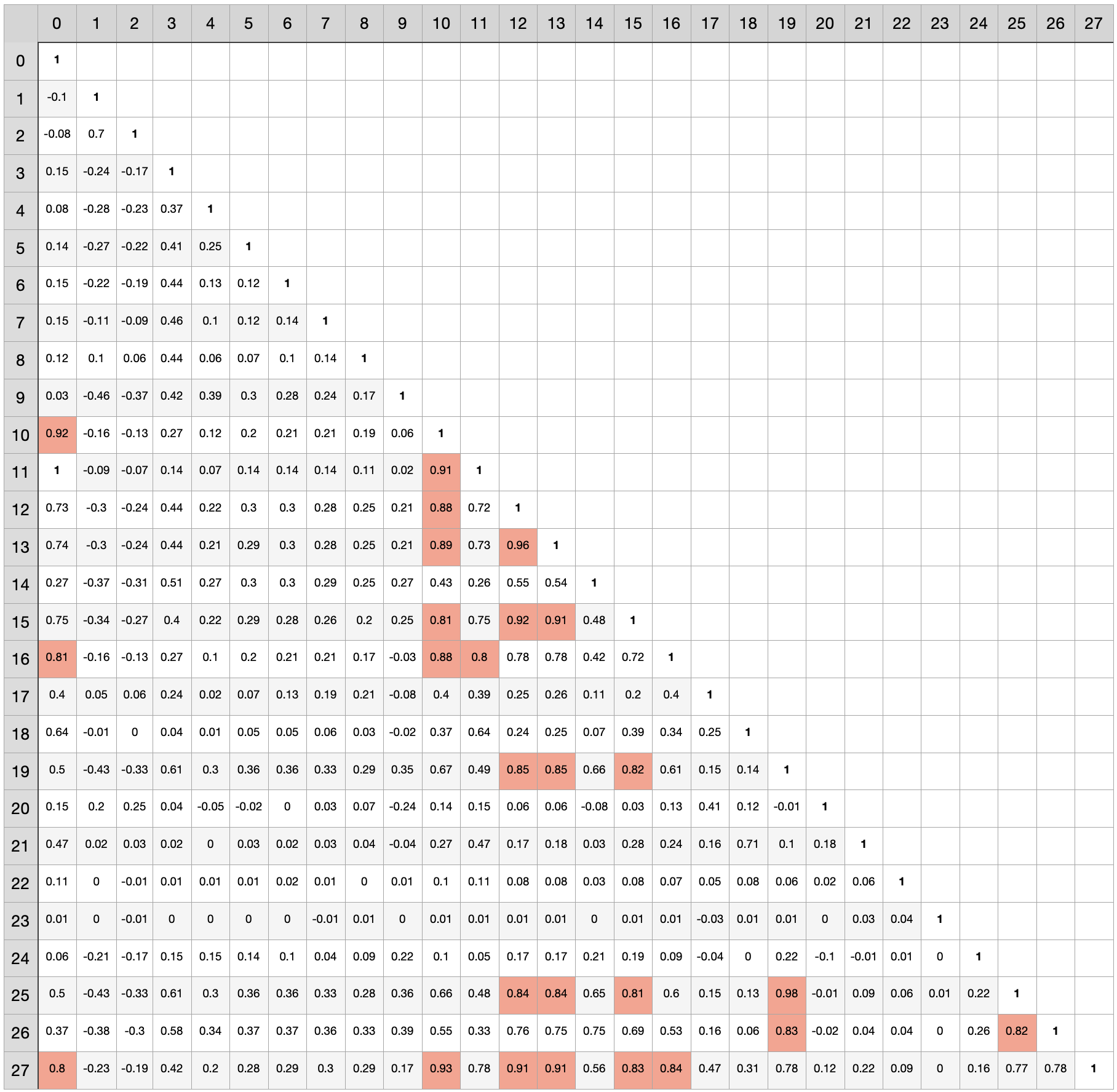}
     \caption{\em Pearson correlations between the variables. Values greater than or equal to 0.8 are highlighted.}
        \label{f:corrmatrix}
    \end{center}
\end{figure}

As shown in Table~\ref{table:corr}, variables 23, 21, 22, 17, 18 have very small correlation with the true muon energy (indicated by both Pearson and Spearman coefficients); they are also among the variables with the smallest variance (Table \ref{table:variance}). In Fig.~\ref{f:corrmatrix} it can be seen that variables 10, 12, 13, 15, 16, 19, 25 and 27 are highly correlated with each other. Out of these, 10, 15 and 16 score lower than the others on the criteria in Tables~\ref{table:corr} and~\ref{table:variance}. Therefore, variables 10, 15, 16, 17, 18, 21, 22, 23 are not useful according to these tests. The features that are considered most useful are 9, 14, 19, 25, 26, since they are highly correlated with true muon energy and also have a high variance.

The results of these tests are only partly in accordance with results of the procedures described in Sec.~\ref{s:pruning}, on which we are more reliant because of the considerations made there.

\begin{table}[h]
    \begin{subtable}[h]{0.45\textwidth}
        \centering
 \begin{tabular}{ c | c }
 Variable & Pearson correlation with true energy \\ 
\hline
23 & 0.000752517 \\ 
21 & 0.00847617 \\ 
17 & -0.00990135 \\
22 & 0.0161485 \\
18 & 0.0382927 \\
11 & 0.130248 \\
0 & 0.148816 \\
20 & -0.154216 \\
16 & 0.204585 \\
8 & 0.205045 \\
10 & 0.224604 \\
7 & 0.25805 \\
24 & 0.262993 \\
6 & 0.288012 \\
5 & 0.301154 \\
4 & 0.319215 \\
2 & -0.349797 \\
27 & 0.362985 \\
13 & 0.37686 \\
12 & 0.382955 \\
15 & 0.384839 \\
1 & -0.425254 \\
3 & 0.438138 \\
25 & 0.50712 \\
19 & 0.508464 \\
9 & 0.514739 \\
14 & 0.530933 \\
26 & 0.596103 \\
\end{tabular}
\caption{\em Pearson correlations.}
        \label{table:pearson}
    \end{subtable}
    \hfill
    \begin{subtable}[h]{0.45\textwidth}
        \centering
  \begin{tabular}{ c | c }
 Variable & Spearman correlation with true energy \\ 
\hline
23 & -0.000212693 \\ 
22 & 0.0301219 \\ 
17 & -0.124055 \\
18 & -0.13034 \\
21 & -0.130664 \\
20 & -0.136399 \\
8 & 0.280734 \\
4 & 0.311921 \\
7 & 0.314859 \\
5 & 0.335737 \\
6 & 0.336183 \\
24 & 0.373343 \\
2 & -0.461922 \\
9 & 0.482525 \\
15 & 0.503136 \\
1 & -0.504717 \\
3 & 0.508471 \\
13 & 0.513933 \\
12 & 0.517588 \\
25 & 0.528493 \\
16 & 0.531662 \\
19 & 0.535617 \\
11 & 0.560924 \\
10 & 0.564657 \\
14 & 0.573352 \\
0 & 0.604411 \\
27 & 0.632403 \\
26 & 0.657757 \\
\end{tabular}
\caption{\em Spearman correlations.}
        \label{table:spearman}
     \end{subtable}
     \caption{\em Correlations between each variable and the true energy, in ascending order of absolute value.}
     \label{table:corr}
         \hfill
   
\end{table}

\begin{table}
\begin{center}
\begin{tabular}{ c | c }
 Variable & Variance \\ 
\hline
21 & 2.93677E-05 \\ 
17 & 0.000128188 \\ 
23 & 0.000186787 \\
2 & 0.000257366 \\
22 & 0.000276385 \\
18 & 0.000413712 \\
16 & 0.000662807 \\
3 & 0.00111181 \\
8 & 0.00126114 \\
27 & 0.00131356 \\
11 & 0.00156317 \\
7 & 0.00169358 \\
0 & 0.00176504 \\
1 & 0.00213874 \\
5 & 0.00267488 \\
6 & 0.00273532 \\
15 & 0.00320387 \\
4 & 0.00365186 \\
20 & 0.00470403 \\
14 & 0.00820994 \\
26 & 0.00906648 \\
10 & 0.0098884 \\
12 & 0.0113694 \\
13 & 0.0145185 \\
19 & 0.018794 \\
25 & 0.0198137 \\
9 & 0.0255759 \\
24 & 0.0959122 \\
\end{tabular}
\caption{\em Variance of each variable, computed after normalization, in ascending order.}
        \label{table:variance}
\end{center}
\end{table}

\FloatBarrier


\begin{thebibliography}{99}
\itemsep -2pt 

\bibitem{Linnainmaa_70} S.~Linnainmaa, ``The representation of the cumulative rounding error of an algorithm as a Taylor expansion of the local rounding errors", Univ. Helsinki, 1970.

\bibitem{Werbos_81} P.~Werbos, ``Applications of Advances in Nonlinear Sensitivity Analysis", {\em Proceedings Of The 10th IFIP Conference, 31.8 - 4.9, NYC}, pp. 762-770 (1981).

\bibitem{backprop} D.~Rumelhart, G.~Hinton, and R.~Williams,  ``Learning representations by back-propagating errors", Nature 323 (1986) 533.

\bibitem{jurgen_pretrain} J.~Schmidhuber, ``Learning Complex, Extended Sequences Using the Principle of History Compression", Neural Computation 4 (1992) 234.

\bibitem{hinton_pretrain} G.~Hinton, and R.~Salakhutdinov, ``Reducing the Dimensionality of Data with Neural Networks",  Science 313 (2006) 504, https://science.sciencemag.org/content/313/5786/504.

\bibitem{Glorot} X.~Glorot and Y.~Bengio, ``Understanding the difficulty of training deep feedforward neural networks", JMLR Workshop and Conference Proceedings 9 (2010), http://jmlr.org/proceedings/papers/v9/glorot10a/glorot10a.pdf .

\bibitem{He} K.~He, X.~Zhang, S.~Ren, and J.~Sun, ``Delving Deep into Rectifiers: Surpassing Human-Level Performance on ImageNet Classification", Proc. 2015 IEEE Int. Conf. On Computer Vision (ICCV) (2015) 1026, https://doi.org/10.1109/ICCV.2015.123.

\bibitem{relu} V.~Nair and G.~Hinton, ``Rectified Linear Units Improve Restricted Boltzmann Machines", Proc. 27th Intern. Conf. on Machine Learning (2010) 807.

\bibitem{adagrad} J.~Duchi, E.~Hazan, and Y.~Singer, ``Adaptive Subgradient Methods for Online Learning and Stochastic Optimization", J. Mach. Learn. Res. 12 (2011) 2121, http://dl.acm.org/citation.cfm?id=1953048.2021068.

\bibitem{rmsprop} T.~Tieleman and G.~Hinton, ``Lecture 6.5—RmsProp: Divide the gradient by a running average of its recent magnitude", COURSERA: Neural Networks For Machine Learning (2012).

\bibitem{adam} D.~Kingma and J.~Ba, ``Adam: A Method for Stochastic Optimization", 3rd Intern. Conf. on Learning Representations, ICLR 2015 (2015), http://arxiv.org/abs/1412.6980.

\bibitem{cart} L.~Breiman,  J.~Friedman, R.~Olshen, and C.~ Stone,  ``Classification and regression trees", Wadsworth \& Brooks/Cole Advanced Books \& Software (1984), https://cds.cern.ch/record/2253780.

\bibitem{featsample}
Tin~Kam~Ho, ``Random decision forests'', Proc. 3rd Intern. Conf. on Document Analysis and Recognition, Montreal (1995) 278 doi: 10.1109/ICDAR.1995.598994.

\bibitem{random_forests} L.~Breiman, ``Random Forests", Mach. Learn. 45 (2001) 5, https://doi.org/10.1023/A:1010933404324.

\bibitem{boost}
L.~Breiman, ``Arcing the Edge'',
Ann. Prob. 26 (1998) 1683.

\bibitem{boost2}
J.~H.~Friedman, ``Stochastic gradient boosting'',
Comput. Stat. \& Data Anal. 38 (2002) 367,
doi:10.1016/S0167-9473(01)00065-2.

\bibitem{1d_cnn} Tmp Discussions on: Mechanisms of Action (MoA) Prediction competition. {\em Kaggle}. (2021), https://www.kaggle.com/c/lish-moa/discussion/202256, Accessed: 2022/02/25.

\bibitem{tabnet} S.~Arik and T.~Pfister, ``TabNet: Attentive Interpretable Tabular Learning", Proc. AAAI Conf. on Artificial Intelligence 35, (2021) 6679, https://ojs.aaai.org/index.php/AAAI/article/view/16826.

\bibitem{cat_embed} C.~Guo and F.~Berkhahn, ``Entity Embeddings of Categorical Variables", CoRR. abs/1604.06737 (2016), http://arxiv.org/abs/1604.06737.

\bibitem{muon_disc1} 
Carl~D.~Anderson and Seth~H.~Neddermeyer,
``Cloud Chamber Observations of Cosmic Rays at 4300 Meters Elevation and Near Sea-Level'',
Phys. Rev. 50, 263
doi:10.1103/PhysRev.50.263 .

\bibitem{muon_disc2} 
Seth~H.~Neddermeyer and Carl~D.~Anderson,
``Note on the Nature of Cosmic-Ray Particles'',
Phys. Rev. 51, 884,
doi:10.1103/PhysRev.51.884 .

\bibitem{boostedknn1} W.~Li, Y.~Chen, and Y.~Song, ``Boosted K-nearest neighbor classifiers based on fuzzy granules'', Knowledge-Based Systems 195 (2020), 105606, doi https://doi.org/10.1016/j.knosys.2020.105606.

\bibitem{boostedknn2} P.~Piro,  R.~Nock, W.~ Bel haj ali, F.~ Nielsen, and M.~Barlaud, ``Boosting k-Nearest Neighbors Classification'' (2013), 
isbn = 978-1-4471-5519-5,
doi = 10.1007/978-1-4471-5520-1\_12.

\bibitem{boostedknn3} N.~ García-Pedrajas and D.~ Ortiz-Boyer, ``Boosting k-nearest neighbor classifier by means of input space projection'',
Expert Systems with Applications 36, n.7 (2009) 10570,
doi https://doi.org/10.1016/j.eswa.2009.02.065.

\bibitem{knnhb} T.~ Neo and D.~ Ventura, ``A direct boosting algorithm for the k-nearest neighbor classifier via local warping of the distance metric'', Pattern Recognitio Letters 33 (2012) 92, 
doi = 10.1016/j.patrec.2011.09.028. 


\bibitem{cmsbbb} CMS Collaboration, ``Search for a Higgs boson decaying into a b-quark pair and produced in association with b quarks in proton-proton collisions at 7 TeV", Phys. Lett. B722 (2013) 207, doi: 10.1016/j.physletb.2013.04.017 .

\bibitem{cnnpaper} J.~Kieseler, G.~C.~Strong, F.~Chiandotto, T.~Dorigo, and L.~Layer, ``Calorimetric Measurement of Multi-TeV Muons via Deep Regression", 
Eur. Phys. Journ. C82, 79 (2022), 
https://doi.org/10.1140/epjc/s10052-022-09993-5 .

\bibitem{richter}
J.~E.~Augustin \textit{et al.},
``Discovery of a Narrow Resonance in $e^+ e^-$ Annihilation'',
Phys. Rev. Lett. 33 (1974), 1406-1408
doi:10.1103/PhysRevLett.33.1406

\bibitem{lederman}
S.~W.~Herb \textit{et al.},
``Observation of a Dimuon Resonance at 9.5-GeV in 400-GeV Proton-Nucleus Collisions'',
Phys. Rev. Lett. 39 (1977), 252-255
doi:10.1103/PhysRevLett.39.252

\bibitem{topdisc}
D0 Collaboration, ``Observation of the top quark'',
Phys. Rev. Lett. 74 (1995) 2632,
doi:10.1103/PhysRevLett.74.2632
[arXiv:hep-ex/9503003 [hep-ex]].


\bibitem{rubbia}
UA1 Collaboration,
``Experimental Observation of Isolated Large Transverse Energy Electrons with Associated Missing Energy at s**(1/2) = 540-GeV'',
Phys. Lett. B 122 (1983) 103,
doi:10.1016/0370-2693(83)91177-2.

\bibitem{higgsatlas}
ATLAS Collaboration,
``Observation of a new particle in the search for the Standard Model Higgs boson with the ATLAS detector at the LHC'',
Phys. Lett. B716 (2012) 1,
doi:10.1016/j.physletb.2012.08.020
[arXiv:1207.7214 [hep-ex]].

\bibitem{higgscms}
CMS Collaboration,
``Observation of a New Boson at a Mass of 125 GeV with the CMS Experiment at the LHC'',
Phys. Lett. B716 (2012) 30,
doi:10.1016/j.physletb.2012.08.021
[arXiv:1207.7235 [hep-ex]].

\bibitem{hmmcms}
CMS Collaboration,
``Measurement of Higgs boson decay to a pair of muons in proton-proton collisions at $\sqrt{s}=13\,\mathrm{TeV}$",
CMS-PAS-HIG-19-006, https://cds.cern.ch/record/2725423.

\bibitem{zprime1} 
P.~Fayet, ``Extra U(1)’s and New Forces", Nucl.Phys.B 347 (1990) 743, doi: 10.1016/0550-3213(90)90381-M. 

\bibitem{zprime2} 
P.~Langacker, ``The Physics of Heavy Z0 Gauge Bosons", Rev. Mod. Phys. 81 (2009) 1199, arXiv:0801.1345 [hep-ph], doi: http://dx.doi.org/10.1103/RevModPhys.81.1199.

\bibitem{cmsmuon} 
A.M.~Sirunyan \textit{et al.},
``Performance of the reconstruction and identification of high-momentum muons in proton-proton collisions at $\sqrt{s}$ = 13 TeV'',
Journ. Inst. 15 (2019),
doi:10.1088/1748-0221/15/02/p02027.

\bibitem{atlasmuon}
G.~Aad \textit{et al.},
``Muon reconstruction performance of the ATLAS detector in proton–proton collision data at $\sqrt{s}$ =13 TeV'',
Eur. Phys. J. C76 (2016) 5, 292,
doi:10.1140/epjc/s10052-016-4120-y
[arXiv:1603.05598 [hep-ex]].

\bibitem{pdg}
M.~Tanabashi \textit{et al.} [Particle Data Group],
``Review of Particle Physics'',
Phys. Rev. D 98 (2018) 030001,
doi:10.1103/PhysRevD.98.030001.

\bibitem{uni-approx}
K.~Hornik, M.~Tinchcombe, and H.~White, ``Multilayer Feedforward Networks are Universal Approximators". Neural Networks. Vol. 2. Pergamon Press. pp. 359–366. (1989),
http://cognitivemedium.com/magic\_paper/assets/Hornik.pdf.

\bibitem{GEANT4} 
S.~Agostinelli \textit{et al.} [GEANT4],
``GEANT4--a simulation toolkit",
Nucl. Instrum. Meth. A506 (2003) 250,
doi:10.1016/S0168-9002(03)01368-8.

\bibitem{dataset} J.~Kieseler, G.~C.~Strong, F.~Chiandotto, T.~ Dorigo, and L.~Layer, Preprocessed Dataset for “Calorimetric Measurement of Multi-TeV Muons via Deep Regression" (Zenodo,2021,8), https://doi.org/10.5281/zenodo.5163817.

\bibitem{knn}
N.~Altman, ``An Introduction to Kernel and Nearest-Neighbor Nonparametric Regression", The American Statistician, 46(3), (1992) 175, doi:10.2307/2685209.

\bibitem{hbb}
CMS Collaboration,
``Search for a Higgs Boson Decaying into a b-Quark Pair and Produced in Association with b Quarks in Proton–Proton Collisions at 7 TeV",
Phys. Lett. B722 (2013), 207,
doi:10.1016/j.physletb.2013.04.017.

\bibitem{eusupp}
``2020 Update of the European Strategy for Particle Physics (Brochure)'' (2020), https://cds.cern.ch/record/272137.

\end{thebibliography}
\end{document}